\documentclass{article}%
\usepackage{amsmath}
\usepackage{amsfonts}
\usepackage{amssymb}
\usepackage{graphicx}%
\newcommand{\bpartial}{\mathop{\partial\kern -4pt\raisebox{.8pt}{$|$}}}
\newcommand{\bra}{\mathopen{[\kern-1.6pt[}}
\newcommand{\ket}{\mathclose{]\kern-1.5pt]}}
\newcommand{\bbra}{\mathopen{[\kern-2.2pt[\kern-2.3pt[}}
\newcommand{\bket}{\mathclose{]\kern-2.1pt]\kern-2.3pt]}}

\makeindex
\begin{document}

\title {\large{\bf CLASSIFICATION OF SIX-DIMENSIONAL REAL NILPOTENT LIE BIALGEBRAS OF SYMPLECTIC TYPE AND THEIR POISSON-LIE GROUPS}}
\vspace{3mm}
\author { \small{ \bf A. Poursistani }\hspace{-2mm}{ \footnote{ e-mail: amir.poursistani@azaruniv.ac.ir }}, \small{ \bf  Gh. Haghighatdoost }\hspace{-1mm}{ \footnote{ e-mail: gorbanali@azaruniv.edu - Corresponding author.}}, \small{ \bf J.
Abedi-Fardad }\hspace{-1mm}{\footnote{ e-mail: abedifardadjafar@yahoo.com}} \\
{\small{\em Department of Mathematics,
Azarbaijan Shahid Madani University, Tabriz, Iran}}\\ }
 \maketitle

\begin{abstract}
 In this paper, we classify all six-dimensional real nilpotent Lie bialgebras of symplectic type.
  The Poisson structures on all of the related six-dimensional
Poisson-Lie groups are obtained. Some new
integrable Hamiltonian systems for which the Poisson-Lie group plays the role of a phase space and its dual Lie group plays the role of a symmetry group of the system are obtained.
\end{abstract}

\smallskip

{\bf Keywords:} Lie bialgebra, Poisson-Lie group, Integrable systems.

\section {\large {\bf Introduction}}

The subject of classical integrable systems is closely connected to
the geometry and to the representation theory of Poisson-Lie groups,
together with their associated Lie bialgebras \cite{Drin}, \cite{VGD}. In recent years, non-semisimple Lie
algebras have proved to be significant in various physical applications. Several works
have addressed the classification of low-dimensional
non-semisimple Lie bialgebras
[\cite{JMF} - \cite{RHR}] as well as Lie superbialgebras \cite{ER1}. In this paper, we will try to classify six-dimensional
real nilpotent Lie bialgebras of symplectic type such that on the Lie algebras ${\bf g}$ and their duals ${\tilde {\bf g}}$ we have symplectic structures. The motivation for this classification is that
we are interested in constructing physical models in which the Lie group $G$ (of the Lie algebra ${\bf g}$) plays the role of phase space and
the dual Lie group ${\tilde{G}}$ (of the dual Lie algebra ${\tilde {\bf g}}$) plays the role of symmetry group of the systems (or vice versa).
 The outline of the paper is as follows. In Section 2, we recall the necessary definitions and notations.
  In Section 3, after presenting the list of six-dimensional real Lie algebras of symplectic type \cite{GO}, \cite{MS} based on \cite{JP}
 which provides the classification of real four-dimensional Lie algebras, we classify six-dimensional real nilpotent Lie bialgebras of symplectic type by applying the method described in \cite{ER1}. In Section 4, we calculate the Poisson structures on the Poisson-Lie groups via the map
$\pi(g)$ between the Lie subalgebras ${\bf g}$ and ${\tilde{\bf g}}$. Finally, in Section 5, two integrable systems as examples of physical applications are obtained. For one of
   these examples,
  the Lie group ${\bf A}_{6,27}$ plays the role of phase space and the dual Lie group ${\bf A}_{6,27.xiv}$ plays the role of symmetry group of the systems. Some concluding
  remarks are given in Section 6.


\section {\large {\bf Definitions and notations}}

In this section, we recall standard definitions and propositions
of Lie bialgebras \cite{Drin}, \cite{VGD} (see for a review
\cite{YKS}). Let ${\bf g}$ be a finite-dimensional Lie algebra
and ${\bf g}^\ast$ be its dual space with respect to a
non-degenerate canonical pairing $(\cdot, \cdot)$.

\goodbreak

{\bf Definition}: A {\em Lie bialgebra} is a Lie algebra ${\bf g}$ with a skew-symmetric linear map
$\delta: {\bf g} \rightarrow {\bf g} \otimes {\bf g}$ such that:\\
a) $\delta$ is a one-cocycle, i.e.:
\begin{equation}
\delta ([X,Y])=[\delta (X), 1 \otimes Y + Y \otimes 1] + [1 \otimes X + X \otimes 1, \delta(Y)] \qquad \forall X,Y \in {\bf g},
\end{equation}
where $1$ is the identity map on ${\bf g}$.\\
 b) ${\delta}^t: {\bf
g}^\ast \otimes {\bf g}^\ast \rightarrow {\bf g}^\ast$ is a Lie
bracket on ${\bf g}^\ast$:
\begin{equation}\label{BA6}
(\xi \otimes \eta, \delta(X)) = ({\delta}^t(\xi \otimes \eta), X) = ([\xi, \eta]_{{\bf g}^\ast}, X) \qquad \forall X \in {\bf g}; \; \xi, \eta \in {\bf g}^\ast.
\end{equation}
 The Lie bialgebra defined in this way will be denoted by $({\bf g}, {\bf g}^\ast)$ or $({\bf g}, \delta)$.

{\bf Proposition}: \cite{RHR} If there exists an automorphism $A$ of
${\bf g}$ such that
 \begin{equation}\label{BB9}
 \delta^\prime = (A \otimes A) \circ \delta \circ A^{-1},
 \end{equation}
then the one-cocycles $\delta$ and $\delta^\prime$ of the Lie algebra ${\bf g}$ are equivalent. In this case, the
two Lie bialgebras $({\bf g}, \delta)$ and $({\bf g}, \delta^\prime)$ are equivalent.

{\bf Definition}: A Lie bialgebra $({\bf g}, {\bf g}^\ast)$ is called a {\em coboundary} Lie bialgebra if there exists an element
 $r \in {\bf g} \otimes {\bf g}$ such that:
\begin{equation}\label{BA1}
\delta (X) = [1 \otimes X + X \otimes 1, r] \qquad \forall X \in {\bf g}.
\end{equation}

{\bf Definition}: A {\it Manin triple} is a triple of Lie algebras
$({\cal D}, {\bf g}, {\bf \tilde{g}})$ with a non-degenerate
ad-invariant symmetric bilinear form $\langle \cdot, \cdot \rangle$ on ${\cal D}$, so
that

 \hspace{2mm} 1. ${\bf g}$ and ${\bf \tilde{g}}$ are Lie
subalgebras of ${\cal D}$,

 \hspace{2mm} 2. ${\cal D} = {\bf
g} \oplus {\bf \tilde{g}}$ as a vector space,

 \hspace{2mm}
 3. ${\bf g}$ and ${\bf
\tilde{g}}$ are isotropic with respect to $\langle \cdot, \cdot
\rangle$, i.e.,
\begin{equation}\label{BB5}
\langle X_i, X_j \rangle = \langle \tilde{X}^i, \tilde{X}^j \rangle = 0, \hspace{10mm}
\;\langle X_i, \tilde{X}^j \rangle = {\delta^j}_i.
\end{equation}

There is a one-to-one correspondence between Manin triples $({\cal D}, {\bf
g}, {\bf \tilde{g}})$ with ${\bf \tilde{g}} = {\bf g}^\ast$ and Lie
bialgebras $({\bf g}, {\bf g}^\ast)$ \cite{VC}. If we choose the
structure constants of Lie algebras ${\bf g}$ and ${\bf \tilde{g}}$
as follows:
\begin{equation}\label{BA2}
[X_i, X_j] = f_{ij}^{\;\;k} X_k, \hspace{20mm} [\tilde{X}^i, \tilde{X}^j] = {\tilde{f}}^{ij}_{\;\;k} {\tilde{X}^k},
\end{equation}
with them satisfying
the following Jacobi identities:
\begin{equation}\label{BA7}
f_{ij}^{\;\;k} f_{km}^{\;\;n} + f_{ik}^{\;\;n} f_{mj}^{\;\;k} + f_{jk}^{\;\;n} f_{im}^{\;\;k} = 0,
\end{equation}
\begin{equation}\label{BA8}
{\tilde{f}}^{ij}_{\;\;k} \; {\tilde{f}}^{km}_{\;\;n} +
{\tilde{f}}^{im}_{\;\;k} \; {\tilde{f}}^{jk}_{\;\;n}
+{\tilde{f}}^{jm}_{\;\;k} \; {\tilde{f}}^{ki}_{\;\;n} = 0,
\end{equation}
then, ad-invariance of the bilinear form $\langle \cdot, \cdot \rangle$ on ${\cal D} =
{\bf g} \oplus {\bf \tilde{g}}$ implies that \cite{VC}-\cite{YKS}
\begin{equation}
[X_i, \tilde{X}^j] = {\tilde{f}}^{jk}_{\;\;i} X_k +
f_{ki}^{\;\;j} \tilde{X}^k,
\end{equation}
where, using (\ref{BB5}), (\ref{BA2}), and (\ref{BA6}) we obtain
\begin{equation}\label{BA3}
\delta(X_i) = {\tilde{f}}^{jk}_{\;\;i} X_j \otimes
X_k.
\end{equation}
 By applying the above relation in (1), one can obtain the following mixed Jacobi relation:
\begin{equation}\label{BB1}
f_{kl}^{\;\;m} {\tilde{f}}^{ij}_{\;\;m} =
f_{mk}^{\;\;i} {\tilde{f}}^{jm}_{\;\;l} -
f_{ml}^{\;\;i} {\tilde{f}}^{jm}_{\;\;k} -
f_{mk}^{\;\;j} {\tilde{f}}^{im}_{\;\;l} +
f_{ml}^{\;\;j} {\tilde{f}}^{im}_{\;\;k}.
\end{equation}
This relation can also be obtained from the Jacobi identity of
${\cal D}$.
 \vspace{3mm}

\section {\large {\bf Classification of six-dimensional real nilpotent Lie bialgebras of symplectic type}}

In this section, we classify six-dimensional real nilpotent Lie
bialgebras of symplectic type, such that we have symplectic structures on the Lie
algebras ${\bf g}$ and their duals ${\tilde{\bf g}}$. The
four-dimensional real Lie algebras of symplectic type have been classified in
\cite{GO}. Here, we will use the same method which previously has
been considered in the classification of Lie superbialgebras in
\cite{ER1}. Let us first have a short review about symplectic
structures on Lie algebras and six-dimensional real
Lie algebras of symplectic type and then have a short review about the classification
method of low-dimensional real Lie bialgebras.

\subsection {\large {\bf Six-dimensional real nilpotent Lie algebras of symplectic type}}
A symplectic structure $\omega$ on a $2n$-dimensional Lie algebra ${\bf g}$ is defined as a two-form such that
\footnote{Here we use the cohomology of a Lie algebra such that $d$ is an extrinsic
 derivative on the Lie algebra ${\bf g}$ (see, for example, \cite{DAA}).}\hspace{2mm}

1) $\omega$ is closed, i.e., $d\omega = 0$;\hspace{2mm}

2) $\omega$ has maximal rank, that is, $\omega^n$ is a volume form on the corresponding Lie group.
 The list of six-dimensional real Lie algebras with symplectic structure is given in \cite{GO} (see also \cite{MS});
  and we brought it in Table 1 for self-containing of the paper
  \footnote{Note that in Table 1 we use the Patera classification \cite{JP} of six-dimensional real Lie algebras.}
\\

{\small {\bf Table 1}}: {\small
Six-dimensional real nilpotent Lie algebras of symplectic type.}\\
    \begin{tabular}{l l | l l   }
    \hline\hline
{\footnotesize ${\bf g}$ }&{\footnotesize Non-zero commutation relations }&{\footnotesize ${\bf g}$ }&{\footnotesize Non-zero commutation relations } \\ \hline

{\footnotesize$A_{6,1}$}&{\footnotesize$[X_1,X_2]=X_3,\;[X_1,X_3]=X_4,\;[X_1,X_5]=X_6$} &{\footnotesize$A_{6,2}$}&{\footnotesize$[X_1,X_2]=X_3,\;[X_1,X_3]=X_4,\;[X_1,X_4]=X_5,\;[X_1,X_5]=X_6$} \\

{\footnotesize$A_{6,3}$}&{\footnotesize$[X_1,X_2]=X_6,\;[X_1,X_3]=X_4,\;[X_2,X_3]=X_5$} &{\footnotesize$A_{6,4}$}&{\footnotesize$[X_1,X_2]=X_5,\;[X_1,X_3]=X_6,\;[X_2,X_4]=X_6$} \\

{\footnotesize$A_{6,5}$}&{\footnotesize$[X_1,X_3]=X_5,\;[X_1,X_4]=X_6,\;[X_2,X_3]=k X_6$} &{\footnotesize$A_{6,6}$}&{\footnotesize$[X_1,X_2]=X_6,\;[X_1,X_3]=X_4,\;[X_1,X_4]=X_5,\;[X_2,X_3]=X_5$} \\
{\footnotesize$$}&{\footnotesize$[X_2,X_4]=X_5,\; k=\pm 1$} &{\footnotesize$$}&{\footnotesize$$} \\

{\footnotesize$A_{6,7}$}&{\footnotesize$[X_1,X_3]=X_4,\;[X_1,X_4]=X_5,\;[X_2,X_3]=X_6$} &{\footnotesize$A_{6,8}$}&{\footnotesize$[X_1,X_2]=X_3+X_5,\;[X_1,X_3]=X_4,\;[X_2,X_5]=X_6$} \\

{\footnotesize$A_{6,9}$}&{\footnotesize$[X_1,X_2]=X_3,\;[X_1,X_3]=X_4,\;[X_1,X_5]=X_6$} &{\footnotesize$A_{6,10}$}&{\footnotesize$[X_1,X_2]=X_3,\;[X_1,X_3]=X_5,\;[X_1,X_4]=X_6$} \\

&{\footnotesize$[X_2,X_3]=X_6$} &{\footnotesize$$}&{\footnotesize$[X_2,X_3]=kX_6,\;[X_2,X_4]=X_5,\; k=\pm 1$}\\ {\footnotesize$A_{6,11}$}&{\footnotesize$[X_1,X_2]=X_3,\;[X_1,X_3]=X_4,\;[X_1,X_4]=X_5$} &{\footnotesize$A_{6,15}$}&{\footnotesize$[X_1,X_2]=X_3+X_5,\;[X_1,X_3]=X_4,\;[X_1,X_4]=X_6$} \\
&{\footnotesize$[X_2,X_3]=X_6$} &&{\footnotesize$[X_2,X_5]=X_6$}\\
{\footnotesize$A_{6,16}$}&{\footnotesize$[X_1,X_3]=X_4,\;[X_1,X_4]=X_5,\;[X_1,X_5]=X_6$}&{\footnotesize$A_{6,17}$}&{\footnotesize$[X_1,X_2]=X_3,\;[X_1,X_3]=X_4,\;[X_1,X_4]=X_6$} \\

&{\footnotesize$[X_2,X_3]=X_5,\;[X_2,X_4]=X_6$} &{\footnotesize$$}&{\footnotesize$[X_2,X_5]=X_6$} \\

{\footnotesize$A_{6,18}$}&{\footnotesize$[X_1,X_2]=X_3,\;[X_1,X_3]=X_4,\;[X_1,X_4]=X_6$} &{\footnotesize$A_{6,19}$}&{\footnotesize$[X_1,X_2]=X_3,\;[X_1,X_3]=X_4,\;[X_1,X_4]=X_5$} \\

&{\footnotesize$[X_2,X_3]=X_5,\;[X_2,X_5]=kX_6,\; k \neq 0$} &{\footnotesize$$}&{\footnotesize$[X_1,X_5]=X_6,\;[X_2,X_3]=X_6$} \\

{\footnotesize$A_{6,20}$}&{\footnotesize$[X_1,X_2]=X_3,\;[X_1,X_3]=X_4,\;[X_1,X_4]=X_5$} &{\footnotesize$A_{6,24}$}&{\footnotesize$[X_1,X_2]=X_3$} \\

{\footnotesize$$}&{\footnotesize$[X_1,X_5]=X_6,\;[X_2,X_3]=X_5,\;[X_2,X_4]=X_6$} &{\footnotesize$$}&{\footnotesize$$} \\

{\footnotesize$A_{6,25}$}&{\footnotesize$[X_1,X_2]=X_3,\;[X_4,X_5]=X_6$} &{\footnotesize$A_{6,26}$}&{\footnotesize$[X_1,X_2]=X_3,\;[X_1,X_3]=X_4$} \\
{\footnotesize$A_{6,27}$}&{\footnotesize$[X_3,X_5]=X_1,\;[X_4,X_5]=X_2$} &{\footnotesize$A_{6,28}$}&{\footnotesize$[X_2,X_5]=X_1,\;[X_3,X_5]=X_2,\;[X_4,X_5]=X_3$} \\
{\footnotesize$A_{6,31}$}&{\footnotesize$[X_3,X_4]=X_1,\;[X_2,X_5]=X_1,\;[X_3,X_5]=X_2$} &{\footnotesize$A_{6,32}$}&{\footnotesize$[X_3,X_4]=X_1,\;[X_2,X_5]=X_1,\;[X_3,X_5]=X_2$} \\
{\footnotesize$$}&{\footnotesize$$} &{\footnotesize$$}&{\footnotesize$[X_4,X_5]=X_3$} \\

\smallskip \\
\hline\hline
\end{tabular}

\subsection {\large {\bf Review of the classification method for low-dimensional real Lie bialgebras}}
In this section, we review the method of obtaining and classification
of low-dimensional real Lie bialgebras, as applied for the first time
in \cite{ER1} for classification of real low-dimensional Lie
superbialgebras. For this purpose, we use the following adjoint
representation:
$$ ({\cal X}_i)_j^{\;k} = -{f_{ij}}^k, \qquad ({\cal Y}^k)_{ij} = -{f_{ij}}^k, $$
\begin{equation}
(\tilde{\cal X}^i)^j_{\;k} = -{\tilde{f}}^{ij}_{\;\;k}, \qquad (\tilde{\cal Y}_k)^{ij} = -{\tilde{f}}^{ij}_{\;\;k},
\end{equation}
for writing the matrix forms of equations (\ref{BA8}) and (\ref{BB1}) as follows, respectively,
\begin{equation}\label{BB2}
(\tilde{\cal X}^i)^j_{\;k} \tilde{\cal X}^k + \tilde{\cal X}^i \tilde{\cal X}^j - \tilde{\cal X}^j \tilde{\cal X}^i = 0,
\end{equation}
\begin{equation}\label{BB3}
 (\tilde{\cal X}^i)^j_{\;l} {\cal Y}^l = -(\tilde{{\cal X}}^t)^j {\cal Y}^i + {\cal Y}^j \tilde{{\cal X}}^i - {\cal Y}^i \tilde{{\cal X}}^j + (\tilde{{\cal X}}^t)^i {\cal Y}^j.
\end{equation}
Having the structure constants of the Lie algebra ${\bf g}$
$({f_{ij}}^{\;k})$, we solve matrix equations (\ref{BB2}) and
(\ref{BB3}) in order to obtain the structure constants of the dual
Lie algebras $\tilde{\bf g}$ (${\tilde{f}}^{ij}_{\;\;k}$), such
that $({\bf g}, \tilde{\bf g})$ is a Lie bialgebra. By this method we
will classify six-dimensional real Lie bialgebras of symplectic type.
 We will perform this task in the following three steps similar to \cite{ER1}.\\

{\bf Step 1:} Solving equations (\ref{BB2}) and (\ref{BB3}) and determining the Lie algebras ${\bf g}^\prime$
which are isomorphic to the dual solutions $\tilde{\bf g}$.

With the solution of matrix equations (\ref{BB2}) and (\ref{BB3})
for obtaining matrices $\tilde{{\cal X}}^i$, some structure
constants of $\tilde{\bf g}$ are obtained to be zero, some unknown,
and some obtained in terms of each other. In order to know whether
$\tilde{\bf g}$ is one of the Lie algebras of Table 1 or isomorphic to
them, we must use the following isomorphism relation between the
obtained Lie algebras $\tilde{\bf g}$ and one of the Lie algebras of
Table 1, e.g., ${\bf g}^\prime$. Applying the following
transformation for a change of basis of $\tilde{\bf g}$ as follows
\begin{equation}
{\tilde{X}^{\prime \; i}} = C_{\;j}^i {\tilde{X}^{
j}}, \qquad [{\tilde{X}^{\prime \; i}}, {\tilde{X}^{\prime
\;j}}] = {\tilde{f}^{\prime ij}}_{\;\;k} {\tilde{X}^{\prime \;
k}},
\end{equation}
 we have the following matrix equations for isomorphism:
\begin{equation}\label{BB4}
C(C_{\;j}^i {\tilde{X}^{j}}_{(\tilde{\bf g})}) = {\tilde{X}^{i}}_{({\bf g}^\prime)} C.
\end{equation}
Solving (\ref{BB4}) with the condition $\det C \neq 0$, we obtain
some extra conditions on ${\tilde{f}}^{kl}_{(\tilde{\bf g})\;m}$'s
as imposed by (\ref{BB2}) and (\ref{BB3}).

{\bf Step 2:} Calculate the general form of the transformation matrices
$B: {\bf g}^\prime \rightarrow {\bf g}^\prime.i$ such that $({\bf
g}, {\bf g^\prime}.i)$ are Lie bialgebras.

 As the second step, we
transform the Lie bialgebra $({\bf g}, \tilde{\bf g})$ to the Lie bialgebra
$({\bf g}, {\bf g^\prime}.i)$ (where ${\bf g^\prime}.i$ is isomorphic
as a Lie algebra to ${\bf g^\prime}$) with an automorphism of the Lie
algebra ${\bf g}$. As the inner product (\ref{BB5}) is invariant, we have
$A^{-t}: \tilde{\bf g} \rightarrow {\bf g}^\prime.i$
\begin{equation}
{{X^\prime}_{i}} = A_{i}^{\;k} {X}_{k}, \qquad {\tilde{X}^{\prime
j}} = (A^{-t})_{\;l}^{j} \tilde{X}^{l}, \qquad \langle X_i^\prime,
\tilde{X}^{\prime j} \rangle = {\delta^j}_i,
\end{equation}
where $A^{-t}$ is the inverse transpose of the same matrix $A \in \text{Aut}({\bf g})$. Thus, we have the following relation
\begin{equation}\label{BB6}
(A^{-t})^i_{\;k} \tilde{f}^{kl}_{(\tilde{\bf g})\;m} (A^{-t})^j_{\;l} = {f}^{ij}_{({\bf g}^\prime.i)\;n} (A^{-t})^n_{\;m}.
\end{equation}
Now, for derivation of Lie bialgebras $({\bf g}, {\bf g^\prime}.i)$, we must find the Lie algebras ${\bf g^\prime}.i$ or transformations
 $B: {\bf g}^\prime \rightarrow {\bf g}^\prime.i$, such that
\begin{equation}\label{BB7}
B^i_{\;k} {f}^{kl}_{({\bf g}^\prime)\;m} B^j_{\;l} = {f}^{ij}_{({\bf g}^\prime.i)\;n} B^n_{\;m}.
\end{equation}
For this purpose, it is enough to eliminate ${f}^{ij}_{({\bf g}^\prime.i)\;n}$ between (\ref{BB6}) and (\ref{BB7}). Then, we have the
following matrix equation for $B$:
\begin{equation}\label{BB8}
(A^{-t})^i_{\;m} \tilde{\cal X}^{t\;m}_{(\tilde{\bf g})} A^{-1} = (B^t A)^{-1} (B^i_{\;k} {\cal X}^{t\;k}_{({\bf g}^\prime)}) B^t.
\end{equation}
Now, by solving (\ref{BB8}), we obtain the general form of the matrix $B$
with the condition $\det B \neq 0$.

{\bf Step 3:} Calculate and classify the non-equivalent Lie
bialgebras.

 Having solved (\ref{BB8}), we obtain the general form
of the matrix $B$ so that its elements are written in terms of the
elements of matrices $A, C$, and structure constants
$\tilde{f}^{ij}_{(\tilde{\bf g})\;k}$.
 Now after substituting $B$ in (\ref{BB7}), we obtain the structure constants ${f}^{ij}_{({\bf g^\prime}.i)\;n}$ of the Lie algebra
${\bf g^\prime}.i$ in terms of elements of matrices $A$ and $C$ and
structure constants $\tilde{f}^{ij}_{(\tilde{\bf g})\;k}$. Then, it is checked whether it
is possible to equalize the structure constants ${f}^{ij}_{({\bf
g^\prime}.i)\;n}$ with each other and with $\pm 1$ or not, such that $\det B \neq 0$, $\det A \neq 0$, and $\det C \neq 0$. In this
way, we obtain matrices $B_1, B_2$, etc. Note that in obtaining $B_i$'s we impose the condition $B B_i^{-1} \in \text{Aut}^t({\bf g})$.
 If this condition is not satisfied, then we cannot impose it on the structure constants, because $B$ and $B_i$ are not equivalent (see below).\\
Now, using isomorphism of matrices $B_1, B_2$, etc., we can obtain Lie
bialgebras $({\bf g}, {\bf g}^\prime.i)$, $({\bf g}, {\bf g}^\prime
.ii)$, etc. On the other hand, there remains a question of whether these
Lie bialgebras are equivalent. In order to answer this question, we
use the matrix form of relation (\ref{BB9}). Consider the two Lie
bialgebras $({\bf g}, {\bf g}^\prime.i)$, $({\bf g}, {\bf g}^\prime
.ii)$; then using
\begin{equation}
A(X_i)=A_i^{\;j}X_j,
\end{equation}
the relation (\ref{BB9}) will have the following matrix form:
\begin{equation}\label{BB10}
A^t((A^t)^i_{\;k} {\cal X}_{({\bf g}^\prime.i)}^{\;\;k}) = {\cal X}_{({\bf g}^\prime.ii)}^{\;\;i} A^t.
\end{equation}
 On the other hand, the transformation matrix between ${\bf g}^\prime.i$ and ${\bf g}^\prime.ii$ is $B_2 B_1^{-1}$ such that
$B_1: {\bf g}^\prime \rightarrow {\bf g}^\prime.i$ and $B_2: {\bf g}^\prime \rightarrow {\bf g}^\prime.ii$; then we have
\begin{equation}\label{BB11}
(B_2 B_1^{-1}) ((B_2 B_1^{-1})^i_{\;k} {\cal X}_{({\bf g}^\prime.i)}^{\;\;k}) = {\cal X}_{({\bf g}^\prime.ii)}^{\;\;i} (B_2 B_1^{-1}).
\end{equation}
A comparison of (\ref{BB11}) with (\ref{BB10}) reveals that if
$B_2 B_1^{-1} \in \text{Aut}^t({\bf g})$, then the Lie bialgebras $({\bf g}, {\bf
g}^\prime.i)$ and $({\bf g}, {\bf g}^\prime.ii)$ are equivalent. In
this way, we obtain non-equivalent classes of $B_i$'s, and we consider
only one element of each class. Now, we will use this method
for obtaining and classifying all six-dimensional real
nilpotent Lie bialgebras of symplectic type.

\subsection {\large {\bf Classification of six-dimensional real nilpotent Lie bialgebras of symplectic type}}

In the following, we explain the above
method by describing the details of the calculations for obtaining
the symplectic Lie bialgebra $(A_{6,8}, A_{6,27.vi})$.

\smallskip
{\bf {An example:}}

One of the solutions of the Jacobi and mixed Jacobi identities (\ref{BB2}) and (\ref{BB3}) for the Lie algebra $A_{6,8}$ has the following form:
\begin{equation}
{\tilde{f}}^{34}_{\;\;1} = \alpha, \qquad {\tilde{f}}^{46}_{\;\;1} = \beta, \qquad {\tilde{f}}^{46}_{\;\;2} = \gamma.
\end{equation}
where $\alpha, \beta, \gamma$ are arbitrary constants. Now, using (\ref{BB4}) the corresponding Lie algebra ${\tilde{\bf g}}$ is isomorphic to the Lie algebra $A_{6,27.vi}$, with the
following isomorphism matrix
\begin{equation}
C=\left(
\begin{matrix}
c_{11}&0&0&0&0&0  \cr
 c_{21}&c_{22}&0&0&0&0 \cr
 c_{31}&c_{32}& -\dfrac{c_{11} c_{64}\alpha}{c_{22}\gamma}&0&0&c_{36}\cr
 c_{41}&c_{42}&c_{43}&c_{44}&-\dfrac{c_{22}\gamma}{c_{24}}&c_{64}\cr
 c_{51}&c_{52}&0&0&0&c_{56} \cr
 c_{61}&c_{62}&\dfrac{c_{64} (c_{11}\beta+c_{21}\gamma)}{c_{22}\gamma}&c_{64}&0&c_{66}
 \end{matrix}
\right),
\end{equation}
satisfying the condition $\beta = 0$, where $c_{ij} \in \mathbb{R}$ are arbitrary elements of the isomorphism matrices.
 Now, by substituting the above results and using the following automorphism group element of
the Lie algebra $A_{6,8}$, we have

\begin{equation}
A=\left(
\begin{matrix}
a_{11}&0&a_{13}&a_{14}&a_{15}&a_{16} \cr
 0&a_{22}&a_{23}&a_{24}&a_{25}&a_{26} \cr
0&0&a_{11}a_{22}&a_{34}&0&a_{36}\cr
0&0&0&a_{11}^2 a_{22}&0&0\cr
0&0&0&a_{11}a_{23}-a_{34}&a_{11}a_{12}&-a_{15}a_{22}-a_{36}\cr
0&0&0&0&0&a_{11}a_{22}^2

 \end{matrix}
\right),
\end{equation}
where $a_{11} a_{22} \neq 0$ and $a_{ij}$ are arbitrary elements of the $A$
 matrices. From (\ref{BB8}), one can obtain the following form for the matrix $B$

\begin{equation}
B=\left(
\begin{matrix}
 \frac{a_{11}^2 a_{22}^2b_{33} b_{45} }{ \alpha}& \frac{a_{11}^2 a_{22}^2b_{34} b_{45} }{ \alpha}&0&0&0&0  \cr
- \frac{a_{11}^2 a_{36}b_{33} b_{45} }{ \gamma}- \frac{a_{11}^3 a_{22}^2b_{45} b_{63} }{ \gamma}- \frac{a_{11}^3 a_{22}b_{33} b_{45}\beta }{ \alpha\gamma}&- \frac{a_{11}^2 a_{36}b_{33} b_{45} }{ \gamma}- \frac{a_{11}^3 a_{22}^2b_{45} b_{63} }{ \gamma}- \frac{a_{11}^3 a_{22}b_{33} b_{45}\beta }{ \alpha\gamma}&0&0&0&0 \cr
 b_{31}&b_{32}&b_{33}&b_{34}&0&b_{36}\cr
b_{41}&b_{42}&b_{43}&b_{44}&b_{45}&b_{46}\cr
b_{51}&b_{52}&0&0&0&b_{56}\cr
b_{61}&b_{62}&b_{63}&b_{64}&0&b_{66}
 \end{matrix}
\right)
\end{equation}
where $b_{ij}$ are arbitrary elements of the $B$ matrices. Now,
using (\ref{BB7}) we have found the following commutation relations
for the algebra ${\bf g^\prime}.i$
\begin{equation}
[{\tilde X}^3, {\tilde X}^4] = \frac{\alpha}{a_{11}^2 a_{22}^2} {\tilde X}^1, \qquad [{\tilde X}^4, {\tilde X}^6] = \left(\frac{a_{36}\alpha}{a_{11}^3 a_{22}^4} + \frac{\beta}{a_{11}^2 a_{22}^3}\right) {\tilde X}^1 + \frac{\gamma}{a_{11}^3 a_{22}^2} {\tilde X}^2.
\end{equation}
On the other hand, by choosing $a_{36} = \beta = 0$, we have
 \begin{equation}
B_1=\left(
\begin{matrix}
 \dfrac{a_{11}^2 a_{22}^2 b_{33} b_{45}}{\alpha} & \dfrac{a_{11}^2 a_{22}^2 b_{34} b_{45}}{\alpha} & 0 & 0 & 0 & 0  \cr
-\dfrac{a_{11}^3 a_{22}^2 b_{45} b_{63}}{\gamma} & -\dfrac{a_{11}^3 a_{22}^2 b_{45} b_{63}}{\gamma} & 0 & 0 & 0 & 0 \cr
 b_{31}&b_{32}&b_{33}&b_{34}&0&b_{36}\cr
b_{41}&b_{42}&b_{43}&b_{44}&b_{45}&b_{46}\cr
b_{51}&b_{52}&0&0&0&b_{56}\cr
b_{61}&b_{62}&b_{63}&b_{64}&0&b_{66}
 \end{matrix}
\right),
\end{equation}
where, since $B_1 B^{-1} \in \text{Aut}^t({\bf g})$, $B_1$ is equivalent to $B$.
Furthermore, by choosing $\frac{\alpha}{a_{11}^2 a_{22}^2} = \frac{\gamma}{a_{11}^3 a_{22}^2} = q \neq 0$, we have

  \begin{equation}
B_2=\left(
\begin{matrix}
\dfrac{b_{33} b_{45}}{q} & \dfrac{b_{34} b_{45}}{q} & 0 & 0 & 0 & 0  \cr
-\dfrac{b_{45} b_{63}}{q} & -\dfrac{b_{45} b_{63}}{q} & 0 & 0 & 0 & 0 \cr
 b_{31}&b_{32}&b_{33}&b_{34}&0&b_{36}\cr
b_{41}&b_{42}&b_{43}&b_{44}&b_{45}&b_{46}\cr
b_{51}&b_{52}&0&0&0&b_{56}\cr
b_{61}&b_{62}&b_{63}&b_{64}&0&b_{66}
 \end{matrix}
\right),
\end{equation}
where, since $B_2 B_1^{-1} \in \text{Aut}^t({\bf g})$, $B_2$ is equivalent to $B_1$.
 Now, by choosing $q = 1$, we have
   \begin{equation}
B_3=\left(
\begin{matrix}
b_{33} b_{45} & b_{34} b_{45} & 0 & 0 & 0 & 0  \cr
- b_{45} b_{63} & -b_{45} b_{63} & 0 & 0 & 0 & 0 \cr
 b_{31}&b_{32}&b_{33}&b_{34}&0&b_{36}\cr
b_{41}&b_{42}&b_{43}&b_{44}&b_{45}&b_{46}\cr
b_{51}&b_{52}&0&0&0&b_{56}\cr
b_{61}&b_{62}&b_{63}&b_{64}&0&b_{66}
 \end{matrix}
\right),
\end{equation}
where, since $B_3 B_2^{-1} \in \text{Aut}^t({\bf g})$, $B_3$ is equivalent to $B_2$. So, we obtained the Lie bialgebra
 $(A_{6,8}, A_{6,27.vi})$ such that the commutation relations for $A_{6,27.vi}$
 are as follows:
 \begin{equation}
[{\tilde X}^3, {\tilde X}^4] = {\tilde X}^1, \qquad [{\tilde X}^4, {\tilde X}^6] = {\tilde X}^2.
\end{equation}

 Similarly, we use the above method for the classification of the six-dimensional real nilpotent Lie bialgebras of symplectic type
 \footnote{Note that we consider only Lie bialgebras for which there are symplectic structures on the Lie algebra ${\bf g}$
  and their duals $\tilde{\bf g}$, according to Table 1.}.
  The results of such calculations are given in Table 2 .\\

\newpage
\smallskip

{\small {\bf Table 2}}: {\small
Six-dimensional real nilpotent Lie bialgebras of symplectic type.}\\
    \begin{tabular}{l l l l  p{0.15mm} }
    \hline\hline
{\footnotesize ${\bf g}$ }& {\footnotesize $\tilde{\bf g}$}
&{\footnotesize Non-zero commutation relations of
$\tilde{\bf g}$}& {\footnotesize Comments} \\ \hline

\vspace{2mm}

{\footnotesize $A_{6,1} $}&{\footnotesize $6A_1 $}&
\\

&{\footnotesize$A_{6,1.i} $}&{\footnotesize$[{\tilde
X}^2,{\tilde X}^4]= {\tilde X}^1+ {\tilde X}^5,\;\;\;[{\tilde X}^3,{\tilde
X}^4]= {\tilde X}^6,\;\;\;[{\tilde X}^4,{\tilde
X}^6]= {\tilde X}^5$} &\\

\vspace{1mm}

&{\footnotesize$A_{6,4.i} $}&{\footnotesize$[{\tilde
X}^2,{\tilde X}^4]= {\tilde X}^1,\;\;\;[{\tilde X}^3,{\tilde
X}^4]= {\tilde X}^5,\;\;\;[{\tilde X}^3,{\tilde
X}^6]= {\tilde X}^1$} &\\

\vspace{1mm}

&{\footnotesize$A_{6,6.i} $}&{\footnotesize$[{\tilde
X}^2,{\tilde X}^4]= {\tilde X}^1,\;\;\;[{\tilde X}^3,{\tilde
X}^4]=  {\tilde X}^2+ {\tilde X}^5,[{\tilde X}^3,{\tilde
X}^6]= {\tilde X}^1,[{\tilde X}^4,{\tilde
X}^6]= {\tilde X}^5$} &\\

\vspace{1mm}

&{\footnotesize$A_{6,7.i} $}&{\footnotesize$[{\tilde
X}^2,{\tilde X}^4]= {\tilde X}^1+ {\tilde X}^5,\;\;\;[{\tilde X}^3,{\tilde
X}^4]= {\tilde X}^6,\;\;\;[{\tilde X}^3,{\tilde
X}^6]= {\tilde X}^1$} &\\

\vspace{1mm}

&{\footnotesize$A_{6,8.i} $}&{\footnotesize$[{\tilde
X}^2,{\tilde X}^4]= {\tilde X}^1+ {\tilde X}^5,\;\;\;[{\tilde X}^3,{\tilde
X}^4]= {\tilde X}^2+ {\tilde X}^6,\;\;\;[{\tilde X}^3,{\tilde
X}^6]= {\tilde X}^1$} &\\

\vspace{1mm}

&{\footnotesize$A_{6,11.i} $}&{\footnotesize$[{\tilde
X}^2,{\tilde X}^4]= {\tilde X}^5,\;\;\;[{\tilde X}^3,{\tilde
X}^4]=  {\tilde X}^6,[{\tilde X}^3,{\tilde
X}^6]= {\tilde X}^1,[{\tilde X}^4,{\tilde
X}^6]= {\tilde X}^2$} &\\

\vspace{1mm}

&{\footnotesize$A_{6,15.i} $}&{\footnotesize$[{\tilde
X}^2,{\tilde X}^4]= {\tilde X}^5,\;\;\;[{\tilde X}^3,{\tilde
X}^4]= {\tilde X}^2+ {\tilde X}^6,\;\;\;[{\tilde X}^3,{\tilde
X}^6]= {\tilde X}^1,[{\tilde X}^4,{\tilde
X}^5]= {\tilde X}^1$} &\\

\vspace{1mm}

&{\footnotesize$A_{6,19.i} $}&{\footnotesize$[{\tilde
X}^2,{\tilde X}^4]= {\tilde X}^5,\;\;\;[{\tilde X}^3,{\tilde
X}^4]= {\tilde X}^6,\;\;\;[{\tilde X}^3,{\tilde
X}^6]= {\tilde X}^1,[{\tilde X}^4,{\tilde
X}^5]= {\tilde X}^1,[{\tilde X}^4,{\tilde
X}^6]= {\tilde X}^2$} &\\

\vspace{1mm}

&{\footnotesize$A_{6,24.i} $}&{\footnotesize$[{\tilde
X}^3,{\tilde X}^4]= {\tilde X}^1$} &\\

\vspace{1mm}

&{\footnotesize$A_{6,26.i} $}&{\footnotesize$[{\tilde
X}^2,{\tilde X}^4]= {\tilde X}^1,\;\;\;[{\tilde X}^4,{\tilde
X}^3]= {\tilde X}^1,\;\;\;[{\tilde X}^4,{\tilde
X}^6]=- {\tilde X}^1+ {\tilde X}^2+ {\tilde X}^5$} &\\
\vspace{1mm}

&{\footnotesize$A_{6,27.i} $}&{\footnotesize$[{\tilde
X}^2,{\tilde X}^4]= {\tilde X}^1,[{\tilde
X}^3,{\tilde X}^4]= {\tilde X}^5$} &\\
\vspace{1mm}

&{\footnotesize$A_{6,28.i} $}&{\footnotesize$[{\tilde
X}^2,{\tilde X}^4]= {\tilde X}^5,\;\;\;[{\tilde X}^3,{\tilde
X}^4]= {\tilde X}^2+ {\tilde X}^6,\;\;\;[{\tilde X}^4,{\tilde
X}^5]= {\tilde X}^1$} &\\


{\footnotesize $A_{6,2} $}&{\footnotesize $6A_1 $}&
\\

\vspace{1mm}

&{\footnotesize$A_{6,3.i} $}&{\footnotesize$[{\tilde
X}^4,{\tilde X}^5]= {\tilde X}^1,\;\;\;[{\tilde X}^4,{\tilde
X}^6]= {\tilde X}^2,\;\;\;[{\tilde X}^5,{\tilde
X}^6]= {\tilde X}^3$} &\\

\vspace{1mm}

&{\footnotesize$A_{6,11.ii} $}&{\footnotesize$[{\tilde
X}^3,{\tilde X}^6]= {\tilde X}^2,\;\;\;[{\tilde X}^4,{\tilde
X}^5]=  {\tilde X}^1,[{\tilde X}^4,{\tilde
X}^6]= {\tilde X}^3,[{\tilde X}^5,{\tilde
X}^6]= {\tilde X}^4$} &\\

&{\footnotesize$A_{6,24.ii} $}&{\footnotesize$[{\tilde
X}^5,{\tilde X}^6]=- {\tilde X}^1- {\tilde X}^2$} &\\

\vspace{1mm}

&{\footnotesize$A_{6,27.ii} $}&{\footnotesize$[{\tilde
X}^4,{\tilde X}^6]= {\tilde X}^2,[{\tilde
X}^5,{\tilde X}^6]= {\tilde X}^2+{\tilde X}^3$} &\\


{\footnotesize $A_{6,4} $}&{\footnotesize $6A_1 $}&
\\

\vspace{1mm}

&{\footnotesize$A_{6,2.i} $}&{\footnotesize$[{\tilde X}^4,{\tilde
X}^5]=- {\tilde X}^2$} &\\

\vspace{1mm}

&{\footnotesize$A_{6,15.ii} $}&{\footnotesize$[{\tilde
X}^1,{\tilde X}^5]= {\tilde X}^3+{\tilde X}^4,[{\tilde X}^4,{\tilde
X}^5]=- {\tilde X}^3,[{\tilde X}^5,{\tilde
X}^6]= {\tilde X}^1$} &\\
\vspace{1mm}

&{\footnotesize$A_{6,17.i} $}&{\footnotesize$[{\tilde
X}^1,{\tilde X}^5]= {\tilde X}^3+{\tilde X}^4,[{\tilde X}^4,{\tilde
X}^5]= - {\tilde X}^2- {\tilde X}^3+ {\tilde X}^4$} &\\

\vspace{1mm}

&{\footnotesize$A_{6,19.ii} $}&{\footnotesize$[{\tilde
X}^1,{\tilde X}^5]= {\tilde X}^3+{\tilde X}^4,[{\tilde X}^4,{\tilde
X}^5]= - {\tilde X}^2- {\tilde X}^3+ {\tilde X}^4,[{\tilde X}^5,{\tilde
X}^6]= {\tilde X}^1$} &\\

\vspace{1mm}

&{\footnotesize$A_{6,28.ii} $}&{\footnotesize$[{\tilde
X}^1,{\tilde X}^5]= {\tilde X}^3+{\tilde X}^4,[{\tilde X}^4,{\tilde
X}^5]= - {\tilde X}^2+ {\tilde X}^4$} &\\


{\footnotesize $A_{6,5} $}&{\footnotesize $6A_1 $}&
\\

\vspace{1mm}

&{\footnotesize$A_{6,1.ii} $}&{\footnotesize$[{\tilde
X}^5,{\tilde X}^6]= {\tilde X}^4$} &\\

\vspace{1mm}

&{\footnotesize$A_{6,2.ii} $}&{\footnotesize$[{\tilde
X}^5,{\tilde X}^6]= -{\tilde X}^2$} &\\

\vspace{1mm}

&{\footnotesize$A_{6,24.iii} $}&{\footnotesize$[{\tilde
X}^5,{\tilde X}^6]=- {\tilde X}^4$} &\\

\vspace{1mm}

&{\footnotesize$A_{6,25.i} $}&{\footnotesize$[{\tilde
X}^2,{\tilde X}^6]= {\tilde X}^4,[{\tilde X}^5,{\tilde
X}^6]=  {\tilde X}^2$} &\\

\vspace{1mm}

&{\footnotesize$A_{6,26.ii} $}&{\footnotesize$[{\tilde
X}^5,{\tilde X}^6]= {\tilde X}^2$} &\\

\vspace{1mm}

&{\footnotesize$A_{6,27.iii} $}&{\footnotesize$[{\tilde
X}^2,{\tilde X}^6]= {\tilde X}^4$} &\\

\vspace{1mm}

&{\footnotesize$A_{6,28.iii} $}&{\footnotesize$[{\tilde
X}^2,{\tilde X}^6]=- {\tilde X}^4$} &\\


{\footnotesize $A_{6,6} $}&{\footnotesize $6A_1 $}&
\\

\vspace{1mm}

&{\footnotesize$A_{6,3.ii} $}&{\footnotesize$[{\tilde
X}^2,{\tilde X}^5]=2 {\tilde X}^3,\;\;\;[{\tilde X}^2,{\tilde
X}^6]= {\tilde X}^1,\;\;\;[{\tilde X}^5,{\tilde
X}^6]=- {\tilde X}^4$} &\\
\vspace{1mm}

&{\footnotesize$A_{6,6.ii} $}&{\footnotesize$[{\tilde
X}^2,{\tilde X}^5]=-2 {\tilde X}^3,\;\;\;[{\tilde X}^2,{\tilde
X}^6]=- {\tilde X}^1,\;\;\;[{\tilde X}^3,{\tilde
X}^5]={\tilde X}^1,[{\tilde X}^5,{\tilde
X}^6]= {\tilde X}^4$} &\\

\vspace{1mm}

&{\footnotesize$A_{6,8.ii} $}&{\footnotesize$[{\tilde
X}^2,{\tilde X}^5]= {\tilde X}^3,\;\;\;[{\tilde X}^2,{\tilde
X}^6]= {\tilde X}^1,\;\;\;[{\tilde X}^4,{\tilde
X}^6]={\tilde X}^1,[{\tilde X}^5,{\tilde
X}^6]= {\tilde X}^2$} &\\

\vspace{1mm}

&{\footnotesize$A_{6,15.iii} $}&{\footnotesize$[{\tilde
X}^2,{\tilde X}^5]= {\tilde X}^1,\;\;\;[{\tilde X}^2,{\tilde
X}^6]= {\tilde X}^1,\;\;\;[{\tilde X}^3,{\tilde
X}^5]={\tilde X}^1,[{\tilde X}^4,{\tilde
X}^5]={\tilde X}^3,[{\tilde X}^4,{\tilde
X}^6]={\tilde X}^1,[{\tilde X}^5,{\tilde
X}^6]= {\tilde X}^4$} &\\

\vspace{1mm}

&{\footnotesize$A_{6,17.ii} $}&{\footnotesize$[{\tilde
X}^2,{\tilde X}^6]= {\tilde X}^1,\;\;\;[{\tilde X}^3,{\tilde
X}^5]=- {\tilde X}^1,\;\;\;[{\tilde X}^4,{\tilde
X}^5]={\tilde X}^3,[{\tilde X}^5,{\tilde
X}^6]= {\tilde X}^4$} &\\

&{\footnotesize$A_{6,26.iii} $}&{\footnotesize$[{\tilde
X}^2,{\tilde X}^5]=- {\tilde X}^1,\;\;\;[{\tilde X}^2,{\tilde
X}^6]= {\tilde X}^1,\;\;\;[{\tilde X}^4,{\tilde
X}^5]={\tilde X}^1,[{\tilde X}^4,{\tilde
X}^6]=- {\tilde X}^1,[{\tilde X}^5,{\tilde
X}^6]=2 {\tilde X}^2+{\tilde X}^4$} &\\

&{\footnotesize$A_{6,27.iv} $}&{\footnotesize$[{\tilde
X}^2,{\tilde X}^5]=- {\tilde X}^1+ {\tilde X}^3,\;\;\;[{\tilde X}^2,{\tilde
X}^6]= {\tilde X}^1,\;\;\;[{\tilde X}^4,{\tilde
X}^5]={\tilde X}^1-{\tilde X}^3,[{\tilde X}^4,{\tilde
X}^6]=- {\tilde X}^1$} &\\

&{\footnotesize$A_{6,32.i} $}&{\footnotesize$[{\tilde
X}^2,{\tilde X}^5]= {\tilde X}^3,\;\;\;[{\tilde X}^2,{\tilde
X}^6]= {\tilde X}^1,\;\;\;[{\tilde X}^3,{\tilde
X}^5]={\tilde X}^1,[{\tilde X}^5,{\tilde
X}^6]= {\tilde X}^2$} &\\

\hline\hline
\end{tabular}
\newpage

{\small {\bf Table 2}}: {\small
 (Continued.)}\\
    \begin{tabular}{l l l l  p{0.15mm} }
    \hline\hline
{\footnotesize ${\bf g}$ }& {\footnotesize $\tilde{\bf g}$}
&{\footnotesize Non-zero commutation relations of
$\tilde{\bf g}$}& {\footnotesize Comments} \\ \hline
\vspace{1mm}

\vspace{1mm}

{\footnotesize $A_{6,7} $}&{\footnotesize $6A_1 $}&\\

\vspace{1mm}

&{\footnotesize$A_{6,1.iii} $}&{\footnotesize$[{\tilde X}^4,{\tilde
X}^5]= {\tilde X}^6,\;\;\;[{\tilde X}^5,{\tilde
X}^6]= {\tilde X}^2$} &\\

\vspace{1mm}

&{\footnotesize$A_{6,8.iii} $}&{\footnotesize$[{\tilde
X}^2,{\tilde X}^5]= {\tilde X}^1,\;\;\;[{\tilde X}^4,{\tilde
X}^5]= {\tilde X}^1,\;\;\;[{\tilde X}^4,{\tilde
X}^6]=2{\tilde X}^3,[{\tilde X}^5,{\tilde
X}^6]= {\tilde X}^4$} &\\

\vspace{1mm}

&{\footnotesize$A_{6,15.iv} $}&{\footnotesize$[{\tilde X}^2,{\tilde
X}^5]= {\tilde X}^1,\;\;\;[{\tilde X}^4,{\tilde
X}^6]=2{\tilde X}^3,[{\tilde X}^5,{\tilde
X}^6]={\tilde X}^4$} &\\

\vspace{1mm}

&{\footnotesize$A_{6,24.iv} $}&{\footnotesize$[{\tilde
X}^4,{\tilde X}^5]=  {\tilde X}^2+ {\tilde X}^3$} &\\
\vspace{1mm}

&{\footnotesize$A_{6,27.v} $}&{\footnotesize$[{\tilde X}^4,{\tilde
X}^5]= {\tilde X}^3,\;\;\;[{\tilde X}^5,{\tilde
X}^6]= {\tilde X}^2+{\tilde X}^3$} &\\

\vspace{1mm}

&{\footnotesize$A_{6,28.iv} $}&{\footnotesize$[{\tilde X}^2,{\tilde
X}^5]= {\tilde X}^3,\;\;\;[{\tilde X}^4,{\tilde
X}^5]= {\tilde X}^1+{\tilde X}^6$} &\\

\vspace{1mm}

&{\footnotesize$A_{6,32.ii} $}&{\footnotesize$[{\tilde X}^4,{\tilde
X}^5]= {\tilde X}^1,\;\;\;[{\tilde X}^4,{\tilde
X}^6]={\tilde X}^3,[{\tilde X}^5,{\tilde
X}^6]= {\tilde X}^4$} &\\

\vspace{1mm}

{\footnotesize $A_{6,8} $}&{\footnotesize $6A_1 $}&
\\

\vspace{1mm}

&{\footnotesize$A_{6,8.iv} $}&{\footnotesize$[{\tilde
X}^3,{\tilde X}^4]= {\tilde X}^1,\;\;\;[{\tilde X}^3,{\tilde
X}^6]= {\tilde X}^2,\;\;\;[{\tilde X}^4,{\tilde
X}^5]={\tilde X}^1,[{\tilde X}^4,{\tilde
X}^6]= {\tilde X}^3$} &\\

\vspace{1mm}

&{\footnotesize$A_{6,15.v} $}&{\footnotesize$[{\tilde
X}^1,{\tilde X}^4]= {\tilde X}^2,\;\;\;[{\tilde X}^3,{\tilde
X}^6]= {\tilde X}^2,\;\;\;[{\tilde X}^4,{\tilde
X}^5]={\tilde X}^1,[{\tilde X}^4,{\tilde
X}^6]=- {\tilde X}^3+ {\tilde X}^5$} &\\

\vspace{1mm}

&{\footnotesize$A_{6,24.v} $}&{\footnotesize$[{\tilde
X}^4,{\tilde X}^5]= {\tilde X}^2$} &\\

\vspace{1mm}

&{\footnotesize$A_{6,26.iv} $}&{\footnotesize$[{\tilde
X}^3,{\tilde X}^4]= {\tilde X}^2,\;\;\;[{\tilde X}^4,{\tilde
X}^6]=- {\tilde X}^3+ {\tilde X}^5$} &\\

\vspace{1mm}

&{\footnotesize$A_{6,27.vi} $}&{\footnotesize$[{\tilde
X}^3,{\tilde X}^4]= {\tilde X}^1,\;\;\;[{\tilde X}^4,{\tilde
X}^6]= {\tilde X}^2$} &\\

\vspace{1mm}


{\footnotesize $A_{6,9} $}&{\footnotesize $6A_1$}&
\\

\vspace{1mm}

&{\footnotesize$A_{6,3.iii} $}&{\footnotesize$[{\tilde
X}^4,{\tilde X}^5]={\tilde X}^2,\;\;\;[{\tilde X}^4,{\tilde
X}^6]={\tilde X}^3,\;\;\;[{\tilde X}^5,{\tilde
X}^6]= {\tilde X}^1$} &\\

\vspace{1mm}

&{\footnotesize$A_{6,6.iii} $}&{\footnotesize$[{\tilde
X}^3,{\tilde X}^6]=-{\tilde X}^2,\;\;\;[{\tilde X}^4,{\tilde
X}^5]={\tilde X}^2,[{\tilde X}^4,{\tilde
X}^6]={\tilde X}^3,\;\;\;[{\tilde X}^5,{\tilde
X}^6]= {\tilde X}^1$} &\\

\vspace{1mm}

&{\footnotesize$A_{6,8.v} $}&{\footnotesize$[{\tilde
X}^3,{\tilde X}^6]={\tilde X}^1-{\tilde X}^2,\;\;\;[{\tilde X}^4,{\tilde
X}^5]={\tilde X}^2,[{\tilde X}^4,{\tilde
X}^6]={\tilde X}^5,\;\;\;[{\tilde X}^5,{\tilde
X}^6]= {\tilde X}^1+{\tilde X}^2$} &\\

&{\footnotesize$A_{6,9.i} $}&{\footnotesize$[{\tilde
X}^3,{\tilde X}^6]={\tilde X}^2,\;\;\;[{\tilde X}^4,{\tilde
X}^5]=-{\tilde X}^2,[{\tilde X}^4,{\tilde
X}^6]={\tilde X}^2,\;\;\;[{\tilde X}^5,{\tilde
X}^6]= {\tilde X}^1$} &\\

\vspace{1mm}

&{\footnotesize$A_{6,24.vi} $}&{\footnotesize$[{\tilde
X}^4,{\tilde X}^6]= {\tilde X}^1+{\tilde X}^2$} &\\

\vspace{1mm}

&{\footnotesize$A_{6,27.vii} $}&{\footnotesize$[{\tilde
X}^3,{\tilde X}^4]= {\tilde X}^2,\;\;\;[{\tilde X}^3,{\tilde
X}^6]= {\tilde X}^1$} &\\


{\footnotesize $A_{6,10} $}&{\footnotesize $6A_1 $}&
\\

\vspace{1mm}

&{\footnotesize$A_{6,3.iv} $}&{\footnotesize$[{\tilde
X}^4,{\tilde X}^5]={\tilde X}^2,\;\;\;[{\tilde X}^4,{\tilde
X}^6]=-2{\tilde X}^1,\;\;\;[{\tilde X}^5,{\tilde
X}^6]= {\tilde X}^3$} &\\

\vspace{1mm}

&{\footnotesize$A_{6,4.ii} $}&{\footnotesize$[{\tilde
X}^3,{\tilde X}^5]=-{\tilde X}^1,\;\;\;[{\tilde X}^4,{\tilde
X}^5]={\tilde X}^2,\;\;\;[{\tilde X}^4,{\tilde
X}^6]= -2{\tilde X}^1$} &\\

\vspace{1mm}

&{\footnotesize$A_{6,24.vii} $}&{\footnotesize$[{\tilde
X}^4,{\tilde X}^5]= {\tilde X}^1$} &\\

\vspace{1mm}

&{\footnotesize$A_{6,31.i} $}&{\footnotesize$[{\tilde X}^3,{\tilde
X}^6]={\tilde X}^1$} &\\

\vspace{1mm}

&{\footnotesize$A_{6,32.iii} $}&{\footnotesize$[{\tilde X}^4,{\tilde
X}^5]={\tilde X}^2,[{\tilde X}^4,{\tilde
X}^6]=-2{\tilde X}^1,\;\;\;[{\tilde X}^5,{\tilde
X}^6]= {\tilde X}^3-{\tilde X}^4$} &\\


{\footnotesize $A_{6,11} $}&{\footnotesize $4A_1 $}&
\\

\vspace{1mm}

&{\footnotesize$A_{6,24.viii} $}&{\footnotesize$[{\tilde
X}^5,{\tilde X}^6]= {\tilde X}^2$} &\\

\vspace{1mm}

&{\footnotesize$A_{6,26.v} $}&{\footnotesize$[{\tilde
X}^2,{\tilde X}^6]={\tilde X}^1,\;\;\;[{\tilde X}^3,{\tilde
X}^6]=-{\tilde X}^1,[{\tilde X}^4,{\tilde
X}^6]={\tilde X}^1,\;\;\;[{\tilde X}^5,{\tilde
X}^6]= -{\tilde X}^1+{\tilde X}^2$} &\\

\vspace{1mm}

&{\footnotesize$A_{6,27.viii} $}&{\footnotesize$[{\tilde
X}^3,{\tilde X}^4]=- {\tilde X}^1,\;\;\;[{\tilde X}^3,{\tilde
X}^6]= {\tilde X}^2$} &\\


{\footnotesize $A_{6,15} $}&{\footnotesize $6A_1 $}&
\\

\vspace{1mm}

&{\footnotesize$A_{6,3.v} $}&{\footnotesize$[{\tilde X}^4,{\tilde
X}^6]= {\tilde X}^5$} &\\

\vspace{1mm}

&{\footnotesize$A_{6,27.ix} $}&{\footnotesize$[{\tilde X}^3,{\tilde
X}^4]={\tilde X}^6$} &\\

{\footnotesize $A_{6,16} $}&{\footnotesize $6A_1 $}&
\\

\vspace{1mm}

&{\footnotesize$A_{6,3.vi} $}&{\footnotesize$[{\tilde
X}^4,{\tilde X}^5]=-{\tilde X}^1,\;\;\;[{\tilde X}^4,{\tilde
X}^6]={\tilde X}^2,\;\;\;[{\tilde X}^5,{\tilde
X}^6]= {\tilde X}^3$} &\\

\vspace{1mm}

&{\footnotesize$A_{6,4.iii} $}&{\footnotesize$[{\tilde
X}^3,{\tilde X}^6]=-{\tilde X}^2,\;\;\;[{\tilde X}^4,{\tilde
X}^5]={\tilde X}^2,\;\;\;[{\tilde X}^4,{\tilde
X}^6]= {\tilde X}^1$} &\\

\vspace{1mm}

&{\footnotesize$A_{6,6.iv} $}&{\footnotesize$[{\tilde
X}^3,{\tilde X}^6]=2{\tilde X}^1,\;\;\;[{\tilde X}^4,{\tilde
X}^5]={\tilde X}^1,\;\;\;[{\tilde X}^4,{\tilde
X}^6]= {\tilde X}^2,[{\tilde X}^5,{\tilde
X}^6]= {\tilde X}^3$} &\\

\vspace{1mm}

&{\footnotesize$A_{6,8.vi} $}&{\footnotesize$[{\tilde
X}^3,{\tilde X}^5]={\tilde X}^1,\;\;\;[{\tilde X}^3,{\tilde
X}^6]={\tilde X}^1-{\tilde X}^2,\;\;\;[{\tilde X}^4,{\tilde
X}^5]= {\tilde X}^1+{\tilde X}^2,[{\tilde X}^5,{\tilde
X}^6]= {\tilde X}^3$} &\\

\vspace{1mm}

&{\footnotesize$A_{6,24.ix} $}&{\footnotesize$[{\tilde
X}^5,{\tilde X}^6]= {\tilde X}^2$} &\\

\vspace{1mm}

&{\footnotesize$A_{6,27.x} $}&{\footnotesize$[{\tilde
X}^4,{\tilde X}^6]= {\tilde X}^1,[{\tilde
X}^5,{\tilde X}^6]= {\tilde X}^3$} &\\

\hline\hline
\end{tabular}
\newpage
{\small {\bf Table 2}}: {\small
 (Continued.)}\\
    \begin{tabular}{l l l l  p{0.15mm} }
    \hline\hline
{\footnotesize ${\bf g}$ }& {\footnotesize $\tilde{\bf g}$}
&{\footnotesize Non-zero commutation relations of
$\tilde{\bf g}$}& {\footnotesize Comments} \\ \hline

\vspace{1mm}

{\footnotesize $A_{6,17} $}&{\footnotesize $6A_1 $}&
\\

\vspace{1mm}

&{\footnotesize$A_{6,3.vii} $}&{\footnotesize$[{\tilde
X}^4,{\tilde X}^5]={\tilde X}^1,\;\;\;[{\tilde X}^4,{\tilde
X}^6]=-{\tilde X}^3,[{\tilde X}^5,{\tilde
X}^6]={\tilde X}^2$} &\\

\vspace{1mm}

&{\footnotesize$A_{6,8.vii} $}&{\footnotesize$[{\tilde
X}^3,{\tilde X}^4]={\tilde X}^1,\;\;\;[{\tilde X}^3,{\tilde
X}^6]=2{\tilde X}^2,[{\tilde X}^4,{\tilde
X}^5]={\tilde X}^1,[{\tilde X}^4,{\tilde
X}^6]={\tilde X}^3$} &\\

\vspace{1mm}

&{\footnotesize$A_{6,15.vi} $}&{\footnotesize$[{\tilde
X}^2,{\tilde X}^6]={\tilde X}^1,\;\;\;[{\tilde X}^3,{\tilde
X}^4]={\tilde X}^1,[{\tilde X}^3,{\tilde
X}^6]={\tilde X}^2,[{\tilde X}^4,{\tilde
X}^6]={\tilde X}^3$} &\\

\vspace{1mm}

&{\footnotesize$A_{6,17.iii} $}&{\footnotesize$[{\tilde X}^3,{\tilde
X}^4]=-{\tilde X}^1$} &\\

\vspace{1mm}

&{\footnotesize$A_{6,24.x} $}&{\footnotesize$[{\tilde
X}^4,{\tilde X}^6]= {\tilde X}^2$} &\\
\vspace{1mm}

&{\footnotesize$A_{6,26.vi} $}&{\footnotesize$[{\tilde
X}^4,{\tilde X}^6]= {\tilde X}^5,[{\tilde
X}^5,{\tilde X}^6]= {\tilde X}^1+{\tilde X}^2$} &\\

\vspace{1mm}

&{\footnotesize$A_{6,27.xi} $}&{\footnotesize$[{\tilde
X}^3,{\tilde X}^6]= {\tilde X}^1,[{\tilde
X}^4,{\tilde X}^6]= {\tilde X}^2+{\tilde X}^5$} &\\

\vspace{1mm}

&{\footnotesize$A_{6,32.iv} $}&{\footnotesize$[{\tilde X}^3,{\tilde
X}^4]={\tilde X}^1,[{\tilde X}^3,{\tilde
X}^6]={\tilde X}^1+{\tilde X}^2,[{\tilde X}^4,{\tilde
X}^6]={\tilde X}^3+{\tilde X}^5$} &\\


{\footnotesize $A_{6,18} $}&{\footnotesize $6A_1 $}&
\\

\vspace{1mm}

&{\footnotesize$A_{6,18.i} $}&{\footnotesize$[{\tilde X}^4,{\tilde
X}^5]={\tilde X}^3$} &\\


{\footnotesize $A_{6,19} $}&{\footnotesize $6A_1 $}&
\\

\vspace{1mm}

&{\footnotesize$A_{6,24.xi} $}&{\footnotesize$[{\tilde
X}^5,{\tilde X}^6]={\tilde X}^1+ {\tilde X}^2$} &\\

\vspace{1mm}

&{\footnotesize$A_{6,26.vii} $}&{\footnotesize$[{\tilde
X}^2,{\tilde X}^5]={\tilde X}^1,[{\tilde X}^3,{\tilde
X}^5]=-{\tilde X}^1,[{\tilde X}^2,{\tilde
X}^6]=-{\tilde X}^1,[{\tilde X}^3,{\tilde
X}^6]=-{\tilde X}^1,[{\tilde X}^4,{\tilde
X}^5]={\tilde X}^1,[{\tilde X}^4,{\tilde
X}^6]=-{\tilde X}^1,$} &\\

\vspace{1mm}

&{\footnotesize$$}&{\footnotesize$[{\tilde X}^5,{\tilde
X}^6]={\tilde X}^1+{\tilde X}^2-{\tilde X}^3-{\tilde X}^4$} &\\

\vspace{1mm}

&{\footnotesize$A_{6,27.xii} $}&{\footnotesize$[{\tilde X}^2,{\tilde
X}^5]={\tilde X}^1,\; [{\tilde X}^2,{\tilde X}^6]=-{\tilde
X}^1,\; [{\tilde X}^3,{\tilde X}^5]=- {\tilde
X}^1, [{\tilde X}^3,{\tilde X}^6]= {\tilde
X}^1,[{\tilde X}^5,{\tilde
X}^6]=-{\tilde X}^2-{\tilde X}^3-{\tilde X}^4$} &\\

\vspace{1mm}


{\footnotesize $A_{6,20} $}&{\footnotesize $6A_1 $}&
\\

\vspace{1mm}

&{\footnotesize$A_{6,17.iv} $}&{\footnotesize$ [{\tilde X}^3,{\tilde X}^5]={\tilde
X}^1,\; [{\tilde X}^3,{\tilde X}^6]= {\tilde
X}^1, [{\tilde X}^4,{\tilde X}^6]= {\tilde
X}^2,[{\tilde X}^5,{\tilde
X}^6]={\tilde X}^4$} &\\

\vspace{1mm}

&{\footnotesize$A_{6,24.xi} $}&{\footnotesize$[{\tilde
X}^3,{\tilde X}^6]={\tilde X}^1+ {\tilde X}^2$} &\\

\vspace{1mm}

&{\footnotesize$A_{6,32.v} $}&{\footnotesize$ [{\tilde X}^3,{\tilde X}^6]=2{\tilde
X}^1,\; [{\tilde X}^4,{\tilde X}^5]= {\tilde
X}^1, [{\tilde X}^4,{\tilde X}^6]= {\tilde
X}^2,[{\tilde X}^5,{\tilde
X}^6]={\tilde X}^4$} &\\


{\footnotesize $A_{6,26} $}&{\footnotesize $6A_1 $}&
\\

\vspace{1mm}

&{\footnotesize$A_{6,24.xii} $}&{\footnotesize$[{\tilde
X}^4,{\tilde X}^5]={\tilde X}^2,[{\tilde
X}^4,{\tilde X}^6]=-{\tilde X}^2$} &\\

\vspace{1mm}

&{\footnotesize$A_{6,26.viii} $}&{\footnotesize$[{\tilde
X}^3,{\tilde X}^4]={\tilde X}^5+{\tilde X}^6,[{\tilde
X}^4,{\tilde X}^5]={\tilde X}^2$} &\\

\vspace{1mm}

&{\footnotesize$A_{6,27.xiii} $}&{\footnotesize$[{\tilde
X}^3,{\tilde X}^4]={\tilde X}^2,[{\tilde
X}^4,{\tilde X}^5]={\tilde X}^5+{\tilde X}^6,[{\tilde
X}^4,{\tilde X}^6]=-{\tilde X}^5-{\tilde X}^6$} &\\

\vspace{1mm}

&{\footnotesize$A_{6,28.v} $}&{\footnotesize$[{\tilde
X}^3,{\tilde X}^4]={\tilde X}^6,[{\tilde
X}^4,{\tilde X}^5]=-{\tilde X}^2+{\tilde X}^5+{\tilde X}^6,[{\tilde
X}^4,{\tilde X}^6]=-{\tilde X}^5-{\tilde X}^6$} &\\


{\footnotesize $A_{6,27} $}&{\footnotesize $6A_1 $}&
\\

\vspace{1mm}

&{\footnotesize$A_{6,3.viii} $}&{\footnotesize$[{\tilde
X}^1,{\tilde X}^2]={\tilde X}^6$} &\\

\vspace{1mm}

&{\footnotesize$A_{6,6.v} $}&{\footnotesize$[{\tilde
X}^1,{\tilde X}^6]={\tilde X}^4$} &\\

\vspace{1mm}

&{\footnotesize$A_{6,7.ii} $}&{\footnotesize$[{\tilde
X}^1,{\tilde X}^6]={\tilde X}^3$} &\\

\vspace{1mm}

&{\footnotesize$A_{6,8.viii} $}&{\footnotesize$[{\tilde
X}^1,{\tilde X}^2]={\tilde X}^5$} &\\

\vspace{1mm}

&{\footnotesize$A_{6,9.ii} $}&{\footnotesize$[{\tilde
X}^1,{\tilde X}^2]={\tilde X}^3+{\tilde X}^4+{\tilde X}^5,[{\tilde
X}^1,{\tilde X}^6]={\tilde X}^4$} &\\

\vspace{1mm}

&{\footnotesize$A_{6,25.ii} $}&{\footnotesize$[{\tilde
X}^1,{\tilde X}^6]={\tilde X}^3$} &\\

\vspace{1mm}

&{\footnotesize$A_{6,27.xiv} $}&{\footnotesize$[{\tilde
X}^1,{\tilde X}^2]=-{\tilde X}^3+{\tilde X}^4$} &\\


{\footnotesize $A_{6,31} $}&{\footnotesize $6A_1 $}&
\\

\vspace{1mm}

&{\footnotesize$A_{6,3.ix} $}&{\footnotesize$[{\tilde
X}^1,{\tilde X}^6]={\tilde X}^4$} &\\

\vspace{1mm}

&{\footnotesize$A_{6,4.iv} $}&{\footnotesize$[{\tilde
X}^1,{\tilde X}^4]={\tilde X}^5,[{\tilde
X}^1,{\tilde X}^6]={\tilde X}^3$} &\\

\vspace{1mm}

&{\footnotesize$A_{6,6.vi} $}&{\footnotesize$[{\tilde
X}^1,{\tilde X}^6]={\tilde X}^3,[{\tilde
X}^2,{\tilde X}^6]={\tilde X}^5$} &\\

\vspace{1mm}

&{\footnotesize$A_{6,8.ix} $}&{\footnotesize$[{\tilde
X}^1,{\tilde X}^4]={\tilde X}^3,[{\tilde
X}^1,{\tilde X}^6]={\tilde X}^2+{\tilde X}^4$} &\\

\vspace{1mm}

&{\footnotesize$A_{6,11.iii} $}&{\footnotesize$[{\tilde
X}^1,{\tilde X}^4]={\tilde X}^3,[{\tilde
X}^1,{\tilde X}^6]={\tilde X}^2$} &\\

\vspace{1mm}

&{\footnotesize$A_{6,17.v} $}&{\footnotesize$[{\tilde
X}^2,{\tilde X}^6]={\tilde X}^5$} &\\

\vspace{1mm}

&{\footnotesize$A_{6,24.xiii} $}&{\footnotesize$[{\tilde
X}^1,{\tilde X}^6]={\tilde X}^5$} &\\

\vspace{1mm}

&{\footnotesize$A_{6,27.xv} $}&{\footnotesize$[{\tilde
X}^1,{\tilde X}^6]={\tilde X}^3$} &\\

\hline\hline
\end{tabular}
\newpage

{\small {\bf Table 2}}: {\small
 (Continued.)}\\
    \begin{tabular}{l l l l  p{0.15mm} }
    \hline\hline
{\footnotesize ${\bf g}$ }& {\footnotesize $\tilde{\bf g}$}
&{\footnotesize Non-zero commutation relations of
$\tilde{\bf g}$}& {\footnotesize Comments} \\ \hline

{\footnotesize $A_{6,32} $}&{\footnotesize $6A_1 $}&
\\

\vspace{1mm}

&{\footnotesize$A_{6,15.vii} $}&{\footnotesize$[{\tilde
X}^1,{\tilde X}^6]=-{\tilde X}^5$} &\\

\vspace{1mm}

&{\footnotesize$A_{6,26.ix} $}&{\footnotesize$[{\tilde
X}^1,{\tilde X}^6]={\tilde X}^5$} &\\

\vspace{1mm}

&{\footnotesize$A_{6,27.xvi} $}&{\footnotesize$[{\tilde
X}^1,{\tilde X}^6]={\tilde X}^4$} &\\

\vspace{1mm}

&{\footnotesize$A_{6,32.vi} $}&{\footnotesize$[{\tilde
X}^1,{\tilde X}^6]=-{\tilde X}^4$} &\\
\vspace{1mm}

\smallskip \\
\hline\hline
\end{tabular}
\smallskip


\section{\large {\bf Calculation of Poisson structures on six-dimensional real nilpotent Poisson-Lie groups}}

In order to compute the Poisson structures on six-dimensional real nilpotent Poisson-Lie groups, we employ the Manin triple construction together with the map $\pi(g)$ between the Lie subalgebras $\mathfrak{g}$ and $\tilde{\mathfrak{g}}$. To obtain the corresponding Poisson-Lie group, we use the following relations:
\begin{equation}
g^{-1} X_i g = a(g)_i^{\;j} X_j, \qquad g^{-1} {\tilde{X}}^i g
= b(g)^{ij} X_j + d(g)^i_{\;j} {\tilde{X}}^j,
\end{equation}
where $a(g)$ and $b(g)$ are the coefficient matrices arising from the adjoint action of $g$ on the Lie algebras $\mathfrak{g}$ and $\tilde{\mathfrak{g}}$. The algebraic Poisson structure is then given by
\begin{equation}
\pi (g) = b(g) a^{-1}(g),
\end{equation}
and the corresponding Poisson structure on the Lie group $G$ can be expressed as
\begin{equation}
P_{G} (g) = (- b(g) a(g)^{-1})^{ij} X_i^{R} \wedge X_j^{R}.
\end{equation}
Equivalently, in terms of tensor components, we have
\begin{equation}\label{BC2}
P^{kl} = (- b(g) a(g)^{-1})^{ij} X_i^{R\;k} X_j^{R\;l}.
\end{equation}

Now, using the right-invariant vector fields as given in \cite{JAF1} and computing the adjoint matrices $a(g)$ and $b(g)$, we obtain the Poisson brackets. The results are summarized in Table 3.
 
\newpage
{\small {\bf Table 3}}: {\small
Poisson brackets over the Poisson-Lie group $G$ (using $\pi(g)$).}\\
    \begin{tabular}{l l l  p{0.15mm} }
    \hline\hline
{\footnotesize ${\bf g}$ }& {\footnotesize $\tilde{\bf g}$}
&{\footnotesize Non-zero Poisson brackets relations } \\
 \hline

{\footnotesize $A_{6,1} $}&{\footnotesize$A_{6,1.i} $}&{\footnotesize$\{x_2,x_4\}=x_1+x_5,\;\{x_3,x_4\}=\frac{3x_1^2}{2}+2x_1x_5+x_6,\;\{x_4,x_6\}=x_5$} \\

\vspace{0.5mm}

&{\footnotesize$A_{6,4.i} $}&{\footnotesize$ \{x_2,x_4\}=qx_1,\;\;\{x_3,x_4\}=\frac{3qx_1^2}{2}+qx_5,\;\;
\{x_3,x_6\}=qx_1,\;\;\{x_4,x_6\}=\frac{3qx_1^2}{2}$} \\

\vspace{0.5mm}

&{\footnotesize$A_{6,6.i} $}&{\footnotesize$ \{x_2,x_4\}=x_1,\;\;\{x_3,x_4\}=\frac{3x_1^2}{2}+x_2+x_5,\;\;
\{x_3,x_6\}=x_1,\;\;\{x_4,x_6\}=\frac{3x_1^2}{2}+x_5$} \\
\vspace{0.5mm}

\vspace{0.5mm}

&{\footnotesize$A_{6,7.i}$}&{\footnotesize$\{x_2,x_4\}=qx_1+qx_5,\;\{x_3,x_4\}=\frac{3qx_1^2}{2}+2qx_1x_5+qx_6,\;\{x_3,x_6\}=qx_1,\;\;\{x_4,x_6\}=\frac{3qx_1^2}{2}$} \\


\vspace{0.5mm}


{\footnotesize $ $}&{\footnotesize$A_{6,8.i} $}&{\footnotesize$ \{x_2,x_4\}=x_1+x_5,\;\{x_3,x_4\}=\frac{3x_1^2}{2}+x_2+2x_1x_5+x_6,\;\{x_3,x_6\}=x_1,\;\;\{x_4,x_6\}=\frac{3x_1^2}{2}$} \\

\vspace{0.5mm}

&{\footnotesize$A_{6,11.i} $}&{\footnotesize$\{x_2,x_4\}=x_5,\;\{x_3,x_4\}=2x_1x_5+x_6,\;\{x_3,x_6\}=x_1,\;\;\{x_4,x_6\}=\frac{3x_1^2}{2}+x_2$} \\

\vspace{0.5mm}

{\footnotesize $ $}&{\footnotesize$A_{6,15.i} $}&{\footnotesize$ \{x_2,x_4\}=x_5,\;\{x_3,x_4\}=x_2+2x_1x_5+x_6,\;\{x_3,x_6\}=x_1,\;\{x_4,x_5\}=x_1,\;\;\{x_4,x_6\}=3x_1^2$} \\

\vspace{0.5mm}

{\footnotesize $ $}&{\footnotesize$A_{6,19.i} $}&{\footnotesize$ \{x_2,x_4\}=x_5,\;\{x_3,x_4\}=2x_1x_5+x_6,\;\{x_3,x_6\}=x_1,\;\{x_4,x_5\}=x_1,\;\;\{x_4,x_6\}=3x_1^2+x_2$} \\

\vspace{0.5mm}

{\footnotesize $ $}&{\footnotesize$A_{6,24.i} $}&{\footnotesize$ \{x_3,x_4\}=x_1$} \\
\vspace{0.5mm}
&{\footnotesize$A_{6,26.i}$}&{\footnotesize$ \{x_2,x_4\}=qx_1,\;\;\;\{x_3,x_4\}=-qx_1+\frac{3qx_1^2}{2},\;\;\;\{x_4,x_6\}=-qx_1+qx_2+qx_5$} \\

\vspace{0.5mm}

&{\footnotesize$A_{6,27.i} $}&{\footnotesize$\{x_2,x_4\}=x_1,\;\;\;\{x_3,x_4\}=\frac{3x_1^2}{2}+x_5$} \\

\vspace{0.5mm}

&{\footnotesize$A_{6,28.i} $}&{\footnotesize$\{x_2,x_4\}=x_5,\;\;\;\{x_3,x_4\}=x_2+2x_1x_5+x_6,\;\;\;\{x_4,x_5\}=x_1,\;\;\;\{x_4,x_6\}=\frac{3x_1^2}{2}$} \\

\vspace{0.5mm}

{\footnotesize $A_{6,2} $}&{\footnotesize$A_{6,3.i} $}&{\footnotesize$\{x_2,x_4\}=x_1+x_5,\;\{x_3,x_4\}=\frac{3x_1^2}{2}+2x_1x_5+x_6,\;\;\;\{x_4,x_6\}=x_5$} \\

\vspace{0.5mm}

&{\footnotesize$A_{6,11.ii} $}&{\footnotesize$\{x_3,x_6\}=qx_2,\;\;\;\{x_4,x_5\}=qx_1,\;\{x_4,x_6\}=\frac{3qx_1^2}{2}+2qx_1x_2+qx_3,\;\{x_5,x_6\}=\frac{7x_1^3}{6}+2qx_1^2x_2+2qx_1x_3+qx_4$} \\

&{\footnotesize $A_{6,24.ii}$}&{\footnotesize$\{x_5,x_6\}=-x_1-x_2$}\\

\vspace{0.5mm}

&{\footnotesize$A_{6,27.ii}$}&{\footnotesize$\{x_4,x_6\}=x_2,\;\;\;\{x_5,x_6\}=x_2+2x_1x_2+x_3$} \\
\vspace{0.5mm}
{\footnotesize $A_{6,4} $}&{\footnotesize $A_{6,2.i} $}&{\footnotesize$\{x_4,x_5\}=-x_2,\;\;\{x_5,x_6\}=\frac{3x_1^2}{2}$} \\

\vspace{0.5mm}

\vspace{0.5mm}
{\footnotesize $$}&{\footnotesize $A_{6,15.ii}$}&{\footnotesize$ \{x_1,x_5\}=x_3+x_4,\;\{x_4,x_5\}=-x_3,\;\{x_5,x_6\}=x_1+2x_2x_3+\frac{x_3^2}{2}+x_3x_4$} \\

\vspace{0.5mm}
{\footnotesize $$}&{\footnotesize $A_{6,17.i}$}&{\footnotesize$ \{x_1,x_5\}=x_3+x_4,\;\{x_4,x_5\}=-x_2-x_3+x_4,\;\{x_5,x_6\}=\frac{3x_2^2}{2}+2x_2x_3+\frac{x_3^2}{2}-2x_2x_4+x_3x_4$} \\

\vspace{0.5mm}
{\footnotesize $$}&{\footnotesize $A_{6,19.ii}$}&{\footnotesize$ \{x_1,x_5\}=x_3+x_4,\;\{x_4,x_5\}=-x_2-x_3+x_4,\;\{x_5,x_6\}=x_1+\frac{3x_2^2}{2}+2x_2x_3+\frac{x_3^2}{2}-2x_2x_4+x_3x_4$} \\

\vspace{0.5mm}
{\footnotesize $$}&{\footnotesize $A_{6,28.ii}$}&{\footnotesize$ \{x_1,x_5\}=x_3+x_4,\;\{x_4,x_5\}=-x_2+x_4,\;\{x_5,x_6\}=\frac{3x_2^2}{2}+\frac{x_3^2}{2}-2x_2x_4+x_3x_4$} \\

\vspace{0.5mm}
{\footnotesize $A_{6,5} $}&{\footnotesize $A_{6,1.ii} $}&{\footnotesize$\{x_5,x_6\}=x_4$} \\

\vspace{0.5mm}

{\footnotesize $ $}&{\footnotesize $A_{6,2.ii}$}&{\footnotesize$ \{x_5,x_6\}=-x_2$} \\

\vspace{0.5mm}

{\footnotesize $ $}&{\footnotesize $A_{6,24.iii}$}&{\footnotesize$ \{x_5,x_6\}=-x_4$} \\

\vspace{0.5mm}

{\footnotesize $ $}&{\footnotesize $A_{6,25.i}$}&{\footnotesize$ \{x_2,x_6\}=x_4,\;\{x_5,x_6\}=x_2-\frac{x_4^2}{2}$} \\

\vspace{0.5mm}

{\footnotesize $ $}&{\footnotesize $A_{6,26.ii}$}&{\footnotesize$ \{x_5,x_6\}=x_2$} \\

\vspace{0.5mm}

{\footnotesize $ $}&{\footnotesize $A_{6,27.iii}$}&{\footnotesize$ \{x_2,x_6\}=x_4,\;\{x_5,x_6\}=-\frac{x_4^2}{2}$} \\

\vspace{0.5mm}

{\footnotesize $ $}&{\footnotesize $A_{6,28.iii}$}&{\footnotesize$ \{x_2,x_6\}=-x_4,\;\{x_5,x_6\}=\frac{x_4^2}{2}$} \\

\vspace{0.5mm}
{\footnotesize $A_{6,6} $}&{\footnotesize $A_{6,3.ii} $}&{\footnotesize$\{x_2,x_5\}=2x_3,\;\{x_2,x_6\}=x_1,\;\{x_5,x_6\}=-4x_1x_3-x_4$} \\

\vspace{0.5mm}
{\footnotesize $ $}&{\footnotesize $A_{6,6.ii} $}&{\footnotesize$\{x_2,x_5\}=-2x_3,\;\{x_2,x_6\}=-x_1,\;\{x_3,x_5\}=x_1,\;\{x_4,x_5\}=\frac{3x_1^2}{2},\;\{x_5,x_6\}=4x_1x_3+x_4$} \\

\vspace{0.5mm}
{\footnotesize $ $}&{\footnotesize $A_{6,8.ii} $}&{\footnotesize$\{x_2,x_5\}=x_3,\;\{x_2,x_6\}=x_1,\;\{x_4,x_6\}=x_1,\;\{x_5,x_6\}=\frac{3x_1^2}{2}+x_2-2x_1x_3$} \\

\vspace{0.5mm}
{\footnotesize $ $}&{\footnotesize $A_{6,15.iii} $}&{\footnotesize$\{x_2,x_5\}=x_1,\;\{x_2,x_6\}=x_1,\;\{x_3,x_5\}=x_1,\;\{x_4,x_6\}=\frac{3x_1^2}{2}+x_3,\;\{x_5,x_6\}=x_4$} \\

\vspace{0.5mm}
{\footnotesize $ $}&{\footnotesize $A_{6,17.ii} $}&{\footnotesize$\{x_2,x_6\}=x_1,\;\{x_3,x_5\}=-x_1,\;\{x_4,x_5\}=-\frac{3x_1^2}{2}+x_3,\;\{x_5,x_6\}=x_4$} \\

\vspace{0.5mm}
{\footnotesize $ $}&{\footnotesize $A_{6,26.iii} $}&{\footnotesize$\{x_2,x_5\}=-x_1,\;\{x_2,x_6\}=x_1,\;\{x_4,x_5\}=x_1,\;\{x_4,x_6\}=-x_1,\;\{x_5,x_6\}=2x_2+x_4$} \\

\vspace{0.5mm}
{\footnotesize $ $}&{\footnotesize $A_{6,27.iv} $}&{\footnotesize$\{x_2,x_5\}=-x_1+x_3,\;\{x_2,x_6\}=x_1,\;\{x_4,x_5\}=x_1-x_3,\;\{x_4,x_6\}=-x_1,\;\{x_5,x_6\}=-2x_1x_3$} \\

\hline\hline
\end{tabular}

\newpage

{\small {\bf Table 3}}: {\small
 (Continued.)}\\
    \begin{tabular}{l l l  p{0.15mm} }
    \hline\hline
{\footnotesize ${\bf g}$ }& {\footnotesize $\tilde{\bf g}$}
&{\footnotesize Non-zero Poisson brackets relations}\\
\hline

\vspace{0.5mm}
{\footnotesize $ $}&{\footnotesize $A_{6,32.i} $}&{\footnotesize$\{x_2,x_5\}=x_3,\;\{x_2,x_6\}=x_1,\;\{x_3,x_5\}=x_1,\;\{x_4,x_5\}=\frac{3x_1^2}{2},\;\{x_5,x_6\}=x_2-2x_1x_3$} \\

\vspace{0.5mm}
{\footnotesize $A_{6,7} $}&{\footnotesize $A_{6,1.iii} $}&{\footnotesize$\{x_4,x_5\}=x_6,\;\{x_5,x_6\}=x_2$} \\

\vspace{0.5mm}
{\footnotesize $ $}&{\footnotesize $A_{6,8.iii} $}&{\footnotesize$\{x_2,x_5\}=x_1,\;\{x_4,x_5\}=x_1,\;\{x_4,x_6\}=2x_3,\;\{x_5,x_6\}=4x_1x_3+x_4$} \\

\vspace{0.5mm}
{\footnotesize $ $}&{\footnotesize $A_{6,15.iv} $}&{\footnotesize$\{x_2,x_5\}=x_1,\;\{x_4,x_6\}=2x_3,\;\{x_5,x_6\}=4x_1x_3+x_4$} \\

\vspace{0.5mm}
{\footnotesize $ $}&{\footnotesize $A_{6,24.iv} $}&{\footnotesize$\{x_4,x_5\}=x_2+x_3$} \\

\vspace{0.5mm}
{\footnotesize $ $}&{\footnotesize $A_{6,27.v} $}&{\footnotesize$\{x_4,x_5\}=x_3,\;\{x_5,x_6\}=x_2+x_3$} \\

\vspace{0.5mm}
{\footnotesize $ $}&{\footnotesize $A_{6,28.iv} $}&{\footnotesize$\{x_2,x_5\}=x_3,\;\{x_4,x_5\}=x_1+x_6,\;\{x_5,x_6\}=\frac{x_3^2}{2}$} \\

\vspace{0.5mm}
{\footnotesize $ $}&{\footnotesize $A_{6,32.ii} $}&{\footnotesize$\{x_4,x_5\}=x_1,\;\{x_4,x_6\}=x_3,\;\{x_5,x_6\}=2x_1x_3+x_4$} \\

\vspace{0.5mm}
{\footnotesize $A_{6,8} $}&{\footnotesize $A_{6,8.iv} $}&{\footnotesize$\{x_3,x_4\}=x_1,\;\{x_3,x_6\}=x_2,\;\{x_4,x_5\}=x_1,\;\{x_4,x_6\}=3x_1x_2+x_3$} \\

\vspace{0.5mm}
{\footnotesize $ $}&{\footnotesize $A_{6,15.v} $}&{\footnotesize$\{x_1,x_4\}=x_2,\;\{x_3,x_4\}=-\frac{x_2^2}{2},\;\{x_3,x_6\}=x_2,\;\{x_4,x_5\}=x_1+\frac{x_2^2}{2},\;\{x_4,x_6\}=3x_1x_2+\frac{2x_2^3}{3}-x_3+x_5$} \\

\vspace{0.5mm}
{\footnotesize $ $}&{\footnotesize $A_{6,24.v} $}&{\footnotesize$\{x_1,x_6\}=1,\;\{x_2,x_4\}=1,\;\{x_3,x_5\}=1,\;\{x_3,x_6\}=x_2,\;\{x_4,x_6\}=x_3-x_5,\;\{x_5,x_6\}=x_2$} \\

\vspace{0.5mm}
{\footnotesize $ $}&{\footnotesize $A_{6,26.iv} $}&{\footnotesize$\{x_3,x_4\}=qx_2,\;\{x_4,x_6\}=-qx_3+qx_5$} \\

\vspace{0.5mm}
{\footnotesize $ $}&{\footnotesize $A_{6,27.vi} $}&{\footnotesize$\{x_3,x_4\}=qx_1,\;\{x_4,x_6\}=qx_2$} \\

\vspace{0.5mm}
{\footnotesize $A_{6,9} $}&{\footnotesize $A_{6,3.iii} $}&{\footnotesize$\{x_4,x_5\}=x_2,\;\{x_4,x_6\}=2x_1x_2+x_3,\;\{x_5,x_6\}=x_1$} \\

\vspace{0.5mm}
{\footnotesize $ $}&{\footnotesize $A_{6,6.iii} $}&{\footnotesize$\{x_3,x_6\}=-x_2,\;\{x_4,x_5\}=x_2,\;\{x_4,x_6\}=x_3,\;\{x_5,x_6\}=x_1$} \\

\vspace{0.5mm}
{\footnotesize $ $}&{\footnotesize $A_{6,8.v} $}&{\footnotesize$\{x_3,x_6\}=x_1-x_2,\;\{x_4,x_5\}=x_2,\;\{x_4,x_6\}=\frac{3x_1^2}{2}+x_5,\;\{x_5,x_6\}=x_1+x_2$} \\

\vspace{0.5mm}
{\footnotesize $ $}&{\footnotesize $A_{6,9.i} $}&{\footnotesize$\{x_3,x_6\}=x_2,\;\{x_4,x_5\}=-x_2,\;\{x_4,x_6\}=x_2,\;\{x_5,x_6\}=x_1$} \\

\vspace{0.5mm}
{\footnotesize $ $}&{\footnotesize $A_{6,24.vi} $}&{\footnotesize$\{x_4,x_6\}=x_1+x_2$} \\

\vspace{0.5mm}
{\footnotesize $ $}&{\footnotesize $A_{6,27.vii} $}&{\footnotesize$\{x_3,x_4\}=x_2,\;\{x_3,x_6\}=x_1,\;\{x_4,x_6\}=\frac{3x_1^2}{2}-\frac{3x_2^2}{2}$} \\

\vspace{0.5mm}
{\footnotesize $A_{6,10} $}&{\footnotesize $A_{6,3.iv} $}&{\footnotesize$\{x_4,x_5\}=x_2,\;\{x_4,x_6\}=-2x_1,\;\{x_5,x_6\}=-4x_1x_2+x_3$} \\

\vspace{0.5mm}
{\footnotesize $ $}&{\footnotesize $A_{6,4.ii} $}&{\footnotesize$\{x_3,x_5\}=-x_1,\;\{x_4,x_5\}=x_2,\;\{x_4,x_6\}=-2x_1,\;\{x_5,x_6\}=-4x_1x_2+kx_1x_2$} \\

\vspace{0.5mm}
{\footnotesize $ $}&{\footnotesize $A_{6,24.vii} $}&{\footnotesize$\{x_4,x_5\}=x_1,\;\{x_5,x_6\}=-\frac{3x_1^2}{2}$} \\

\vspace{0.5mm}
{\footnotesize $ $}&{\footnotesize $A_{6,31.i} $}&{\footnotesize$\{x_3,x_6\}=x_1,\;\{x_5,x_6\}=\frac{3x_1^2}{2}$} \\

\vspace{0.5mm}
{\footnotesize $ $}&{\footnotesize $A_{6,32.iii} $}&{\footnotesize$\{x_4,x_5\}=x_2,\;\{x_4,x_6\}=-2x_1,\;\{x_5,x_6\}=-4x_1x_2+x_3-x_4$} \\

\vspace{0.5mm}
{\footnotesize $A_{6,11} $}&{\footnotesize $A_{6,24.viii} $}&{\footnotesize$\{x_5,x_6\}=x_2$} \\

\vspace{0.5mm}
{\footnotesize $ $}&{\footnotesize $A_{6,26.v} $}&{\footnotesize$\{x_2,x_6\}=x_1,\;\{x_3,x_6\}=-x_1+\frac{3x_1^2}{2},\;\{x_4,x_6\}=x_1-\frac{3x_1^2}{2}+\frac{7x_1^3}{6},$} \\

\vspace{0.5mm}
{\footnotesize $ $}&{\footnotesize $ $}&{\footnotesize$\;\{x_5,x_6\}=-x_1+\frac{3x_1^2}{2}-\frac{7x_1^3}{6}+\frac{5x_1^4}{8}+x_2$} \\

\vspace{0.5mm}
{\footnotesize $ $}&{\footnotesize $A_{6,27.viii} $}&{\footnotesize$\{x_3,x_4\}=-x_1,\;\{x_3,x_5\}=-\frac{3x_1^2}{2},\;\{x_3,x_6\}=x_2,\;\{x_4,x_5\}=-\frac{7x_1^3}{6},\;\{x_4,x_6\}=3x_1x_2$} \\

\vspace{0.5mm}
{\footnotesize $ $}&{\footnotesize $ $}&{\footnotesize$\;\{x_5,x_6\}=\frac{7x_1^2x_2}{2}$} \\

\vspace{0.5mm}
{\footnotesize $A_{6,15} $}&{\footnotesize $A_{6,3.v} $}&{\footnotesize$\{x_4,x_6\}=x_5$} \\

\vspace{0.5mm}
{\footnotesize $ $}&{\footnotesize $A_{6,27.ix} $}&{\footnotesize$\{x_3,x_4\}=x_6,\;\{x_3,x_6\}=2x_1x_6,\;\{x_4,x_6\}=2x_1^2x_6$} \\

\vspace{0.5mm}
{\footnotesize $A_{6,16} $}&{\footnotesize $A_{6,3.vi} $}&{\footnotesize$\{x_4,x_5\}=-x_1,\;\{x_4,x_6\}=-\frac{3x_1^2}{2}+x_2,\;\{x_5,x_6\}=-\frac{7x_1^3}{6}+3x_1x_2+x_3$} \\

\vspace{0.5mm}
{\footnotesize $ $}&{\footnotesize $A_{6,4.iii} $}&{\footnotesize$\{x_3,x_6\}=-x_2,\;\{x_4,x_5\}=x_2,\;\{x_4,x_6\}=x_1,\;\{x_5,x_6\}=\frac{3x_1^2}{2}-3x_2^2$} \\

\vspace{0.5mm}
{\footnotesize $ $}&{\footnotesize $A_{6,6.iv} $}&{\footnotesize$\{x_3,x_6\}=2x_1,\;\{x_4,x_5\}=x_1,\;\{x_4,x_6\}=\frac{9x_1^2}{2}+x_2,\;\{x_5,x_6\}=\frac{7x_1^3}{2}+3x_1x_2$} \\

\vspace{0.5mm}
{\footnotesize $ $}&{\footnotesize $A_{6,8.vi} $}&{\footnotesize$\{x_3,x_5\}=x_1,\;\{x_3,x_6\}=x_1+\frac{3x_1^2}{2}-x_2,\;\{x_4,x_5\}=x_1+\frac{3x_1^2}{2}+x_2,\;\{x_4,x_6\}=\frac{7x_1^3}{3}+3x_1^2$} \\

\vspace{0.5mm}
{\footnotesize $ $}&{\footnotesize $ $}&{\footnotesize$\;\{x_5,x_6\}=\frac{7x_1^3}{3}+\frac{5x_1^4}{4}-3x_2^2+x_3$} \\

\vspace{0.5mm}
{\footnotesize $ $}&{\footnotesize $A_{6,24.ix} $}&{\footnotesize$\{x_5,x_6\}=x_2$} \\

\vspace{0.5mm}
{\footnotesize $ $}&{\footnotesize $A_{6,27.x} $}&{\footnotesize$\{x_4,x_6\}=x_1,\;\{x_5,x_6\}=\frac{3x_1^2}{2}+x_3$} \\

\vspace{0.5mm}
{\footnotesize $A_{6,17} $}&{\footnotesize $A_{6,3.vii} $}&{\footnotesize$\{x_4,x_5\}=x_1,\;\{x_4,x_6\}=x_1x_2-x_3,\;\{x_5,x_6\}=-\frac{3x_1^2}{2}+x_2$} \\

\vspace{0.5mm}
{\footnotesize $ $}&{\footnotesize $A_{6,8.vii} $}&{\footnotesize$\{x_3,x_4\}=x_1,\;\{x_3,x_6\}=\frac{3x_1^2}{2}+2x_2,\;\{x_4,x_5\}=x_1,\;\{x_4,x_6\}=\frac{7x_1^3}{6}+5x_1x_2+x_3$} \\

\hline\hline
\end{tabular}

\newpage

{\small {\bf Table 3}}: {\small
 (Continued.)}\\
    \begin{tabular}{l l l  p{0.15mm} }
    \hline\hline
{\footnotesize ${\bf g}$ }& {\footnotesize $\tilde{\bf g}$}
&{\footnotesize Non-zero Poisson brackets relations}\\
\hline

\vspace{0.5mm}

{\footnotesize $ $}&{\footnotesize $ $}&{\footnotesize$\;\{x_5,x_6\}=-\frac{3x_1^2}{2}$} \\

\vspace{0.5mm}
{\footnotesize $ $}&{\footnotesize $A_{6,15.vi} $}&{\footnotesize$\{x_2,x_6\}=x_1,\;\{x_3,x_4\}=x_1,\;\{x_3,x_6\}=3x_1^2+x_2,\;\{x_4,x_6\}=\frac{7x_1^3}{3}+2x_1x_2+x_3$} \\

\vspace{0.5mm}
{\footnotesize $ $}&{\footnotesize $A_{6,17.iii} $}&{\footnotesize$\{x_3,x_4\}=-x_1,\;\{x_3,x_6\}=-\frac{3x_1^2}{2},\;\{x_4,x_6\}=-\frac{7x_1^3}{6}$} \\

\vspace{0.5mm}
{\footnotesize $ $}&{\footnotesize $A_{6,24.x} $}&{\footnotesize$\{x_4,x_6\}=x_2$} \\

\vspace{0.5mm}
{\footnotesize $ $}&{\footnotesize $A_{6,26.vi} $}&{\footnotesize$\{x_4,x_6\}=x_5,\;\{x_5,x_6\}=x_1+x_2$} \\

\vspace{0.5mm}
{\footnotesize $ $}&{\footnotesize $A_{6,27.xi} $}&{\footnotesize$\{x_3,x_6\}=x_1,\;\{x_4,x_6\}=\frac{3x_1^2}{2}+x_2+x_5$} \\

\vspace{0.5mm}
{\footnotesize $ $}&{\footnotesize $A_{6,32.iv} $}&{\footnotesize$\{x_3,x_4\}=x_1,\;\{x_3,x_6\}=x_1+\frac{3x_1^2}{2}+x_2,\;\{x_4,x_6\}=\frac{3x_1^2}{2}+\frac{7x_1^3}{6}+2x_1x_2+x_3+x_5$} \\

\vspace{0.5mm}
{\footnotesize $A_{6,18} $}&{\footnotesize $A_{6,18.i} $}&{\footnotesize$\{x_4,x_5\}=x_3,\;\{x_4,x_6\}=2kx_2x_3,\;\{x_5,x_6\}=-2x_1x_3$} \\

\vspace{0.5mm}
{\footnotesize $A_{6,19} $}&{\footnotesize $A_{6,24.xi} $}&{\footnotesize$\{x_5,x_6\}=x_1+x_2$} \\

\vspace{0.5mm}
{\footnotesize $ $}&{\footnotesize $A_{6,26.vii} $}&{\footnotesize$\{x_2,x_5\}=x_1,\;\{x_2,x_6\}=-x_1+\frac{3x_1^2}{2},\;\{x_3,x_5\}=-x_1+\frac{3x_1^2}{2},\;\{x_3,x_6\}=-x_1-3x_1^2+\frac{7x_1^3}{3}$} \\

\vspace{0.5mm}
{\footnotesize $ $}&{\footnotesize $ $}&{\footnotesize$\;\{x_4,x_5\}=x_1-\frac{3x_1^2}{2}+\frac{7x_1^3}{6},\;\{x_4,x_6\}=-x_1-\frac{7x_1^3}{2}+\frac{15x_1^4}{8},$} \\

\vspace{0.5mm}
{\footnotesize $ $}&{\footnotesize $ $}&{\footnotesize$\;\{x_5,x_6\}=x_1-\frac{3x_1^2}{2}-\frac{15x_1^4}{8}+\frac{31x_1^5}{40}+x_2+x_1x_2-\frac{x_1^2x_2}{2}-x_3-x_4$} \\

\vspace{0.5mm}
{\footnotesize $ $}&{\footnotesize $A_{6,27.xii} $}&{\footnotesize$\{x_2,x_5\}=x_1,\;\{x_2,x_6\}=-x_1+\frac{3x_1^2}{2},\;\{x_3,x_5\}=-x_1+\frac{3x_1^2}{2},\;\{x_3,x_6\}=x_1-3x_1^2+\frac{7x_1^3}{3}$} \\

\vspace{0.5mm}
{\footnotesize $ $}&{\footnotesize $ $}&{\footnotesize$\;\{x_4,x_5\}=-\frac{3x_1^2}{2}+\frac{7x_1^3}{6},\;\{x_4,x_6\}=\frac{3x_1^2}{2}-\frac{7x_1^3}{2}+\frac{15x_1^4}{8},$} \\

\vspace{0.5mm}
{\footnotesize $ $}&{\footnotesize $ $}&{\footnotesize$\;\{x_5,x_6\}=\frac{7x_1^3}{6}-\frac{15x_1^4}{8}+\frac{31x_1^5}{40}-x_2+x_1x_2-\frac{x_1^2x_2}{2}-x_3-x_4$} \\

\vspace{0.5mm}
{\footnotesize $A_{6,20} $}&{\footnotesize $A_{6,17.iv} $}&{\footnotesize$\{x_3,x_5\}=x_1,\;\{x_3,x_6\}=x_1+\frac{3x_1^2}{2},\;\{x_4,x_5\}=\frac{3x_1^2}{2},\;\{x_4,x_6\}=x_2+\frac{3x_1^2}{2}+\frac{7x_1^3}{3}$} \\

\vspace{0.5mm}
{\footnotesize $ $}&{\footnotesize $ $}&{\footnotesize$\;\{x_5,x_6\}=\frac{7x_1^3}{6}+\frac{5x_1^4}{4}+3x_1x_2+x_4$} \\

\vspace{0.5mm}
{\footnotesize $ $}&{\footnotesize $A_{6,24.xi} $}&{\footnotesize$\{x_3,x_6\}=x_1+x_2,\;\{x_4,x_6\}=2x_1x_2,\;\{x_5,x_6\}=\frac{7x_1^3}{6}+x_1x_2+2x_1^2x_2+\frac{3x_2^2}{2}$} \\

\vspace{0.5mm}
{\footnotesize $ $}&{\footnotesize $A_{6,32.v} $}&{\footnotesize$\{x_3,x_6\}=2x_1,\;\{x_4,x_5\}=x_1,\;\{x_4,x_6\}=\frac{9x_1^2}{2}+x_2,\;\{x_5,x_6\}=\frac{7x_1^3}{2}+3x_1x_2+x_4$} \\

\vspace{0.5mm}
{\footnotesize $A_{6,26} $}&{\footnotesize $A_{6,24.xii} $}&{\footnotesize$\{x_4,x_5\}=x_2,\;\{x_4,x_6\}=-x_2$} \\

\vspace{0.5mm}
{\footnotesize $ $}&{\footnotesize $A_{6,26.viii} $}&{\footnotesize$\{x_3,x_4\}=qx_5+qx_6,\;\{x_4,x_5\}=qx_2$} \\

\vspace{0.5mm}
{\footnotesize $ $}&{\footnotesize $A_{6,27.xiii} $}&{\footnotesize$\{x_3,x_4\}=qx_2,\;\{x_4,x_5\}=qx_5+qx_6,\;\{x_4,x_6\}=-qx_5-qx_6$} \\

\vspace{0.5mm}
{\footnotesize $ $}&{\footnotesize $A_{6,28.v} $}&{\footnotesize$\{x_3,x_4\}=qx_6,\;\{x_4,x_5\}=-qx_2+qx_5+qx_6,\;\{x_4,x_6\}=-qx_5-qx_6$} \\

\vspace{0.5mm}
{\footnotesize $A_{6,27} $}&{\footnotesize $A_{6,3.viii} $}&{\footnotesize$\{x_1,x_2\}=x_6$} \\

\vspace{0.5mm}
{\footnotesize $$}&{\footnotesize $A_{6,6.v} $}&{\footnotesize$\{x_1,x_6\}=x_4$} \\

\vspace{0.5mm}
{\footnotesize $$}&{\footnotesize $A_{6,7.ii} $}&{\footnotesize$\{x_1,x_6\}=x_3$} \\

\vspace{0.5mm}
{\footnotesize $$}&{\footnotesize $A_{6,8.viii} $}&{\footnotesize$\{x_1,x_2\}=x_5$} \\

\vspace{0.5mm}
{\footnotesize $$}&{\footnotesize $A_{6,9.ii} $}&{\footnotesize$\{x_1,x_2\}=x_3+x_4+x_5,\;\{x_1,x_6\}=x_4$} \\

\vspace{0.5mm}
{\footnotesize $$}&{\footnotesize $A_{6,25.ii} $}&{\footnotesize$\{x_1,x_6\}=x_3$} \\

\vspace{0.5mm}
{\footnotesize $$}&{\footnotesize $A_{6,27.xiv} $}&{\footnotesize$\{x_1,x_3\}=x_3,\;\{x_2,x_4\}=x_4,\;\{x_5,x_6\}=x_5+x_6$} \\

\vspace{0.5mm}
{\footnotesize $A_{6,31} $}&{\footnotesize $A_{6,3.ix} $}&{\footnotesize$\{x_1,x_6\}=x_4$} \\

\vspace{0.5mm}
{\footnotesize $$}&{\footnotesize $A_{6,4.iv} $}&{\footnotesize$\{x_1,x_4\}=x_5$} \\

\vspace{0.5mm}
{\footnotesize $$}&{\footnotesize $A_{6,6.vi} $}&{\footnotesize$\{x_1,x_6\}=x_3-\frac{x_5^2}{2},\;\{x_2,x_6\}=x_5$} \\

\vspace{0.5mm}
{\footnotesize $$}&{\footnotesize $A_{6,8.ix} $}&{\footnotesize$\{x_1,x_4\}=x_3,\;\{x_1,x_6\}=x_2+x_4$} \\

\vspace{0.5mm}
{\footnotesize $$}&{\footnotesize $A_{6,11.iii} $}&{\footnotesize$\{x_1,x_4\}=x_3,\;\{x_1,x_6\}=x_2$} \\

\vspace{0.5mm}
{\footnotesize $$}&{\footnotesize $A_{6,17.v} $}&{\footnotesize$\{x_1,x_6\}=-\frac{x_5^2}{2},\;\{x_1,x_6\}=x_5$} \\

\vspace{0.5mm}
{\footnotesize $$}&{\footnotesize $A_{6,24.xiii} $}&{\footnotesize$\{x_1,x_6\}=x_5$} \\

\vspace{0.5mm}
{\footnotesize $$}&{\footnotesize $A_{6,27.xv} $}&{\footnotesize$\{x_1,x_6\}=x_3$} \\

\vspace{0.5mm}
{\footnotesize $A_{6,32} $}&{\footnotesize $A_{6,15.vii} $}&{\footnotesize$\{x_1,x_6\}=-x_5$} \\

\vspace{0.5mm}
{\footnotesize $$}&{\footnotesize $A_{6,26.ix} $}&{\footnotesize$\{x_1,x_6\}=x_5$} \\

\vspace{0.5mm}
{\footnotesize $$}&{\footnotesize $A_{6,27.xvi} $}&{\footnotesize$\{x_1,x_6\}=x_4$} \\

\vspace{0.5mm}
{\footnotesize $$}&{\footnotesize $A_{6,32.vi} $}&{\footnotesize$\{x_1,x_6\}=-x_4$} \\

\hline\hline
\end{tabular}

\section{Physical Application: An Example of Integrable Hamiltonian Systems via Symplectic Lie Bialgebras}

In this section, we consider two integrable systems obtained via symplectic Lie bialgebras, where in each case the Lie group \(G\), associated with the Lie bialgebra \((\mathfrak{g}, \tilde{\mathfrak{g}})\), serves as the phase space and its dual Lie group \(\tilde{G}\) acts as the symmetry group of the system, following the formalism introduced in \cite{JAF}.

\subsection*{Example 1: Phase Space \(\mathbf{A}_{6,27}\) with Symmetry Group \(\mathbf{A}_{6,27.xiv}\)}

For this example, the symplectic structure on \(\mathbf{A}_{6,27}\) (see Table~3) is given by the non-zero Poisson brackets
\begin{equation}
\{x_1, x_3\} = x_3, \qquad
\{x_2, x_4\} = x_4, \qquad
\{x_5, x_6\} = x_5 + x_6.
\end{equation}

Applying the procedure of \cite{JAF}, the dynamical functions \(Q_i\) take the form
\begin{align}
Q_1 &= x_3 \ln(x_5 + x_6) - x_5, \\
Q_2 &= -x_4 \ln(x_5 + x_6) + x_5, \\
Q_3 &= x_3, \quad Q_4 = x_4, \quad Q_5 = x_5, \quad Q_6 = c,
\end{align}
where \(c\) is a constant.

These functions satisfy the Poisson brackets
\begin{equation}
\{Q_1, Q_2\} = Q_4 - Q_3, \qquad
\{Q_1, Q_5\} = -Q_3, \qquad
\{Q_2, Q_5\} = Q_4,
\end{equation}
which are equivalently expressed as \(\{Q_i, Q_j\} = f_{ij}^k Q_k\), with \(f_{ij}^k\) being the structure constants of the symmetry Lie algebra \(\mathbf{A}_{6,27.xiv}\). The invariants of the system are either \((Q_1, Q_3, Q_4)\) or \((Q_3, Q_4, Q_5)\), and any one of these \(Q_i\) may be chosen as the Hamiltonian to define an integrable system.

\subsection*{Example 2: Phase Space \(\mathbf{A}_{6,8}\) with Symmetry Group \(\mathbf{A}_{6,24.v}\)}

For this example, the symplectic structure on \(\mathbf{A}_{6,8}\) (see Table~3) is given by
\begin{equation}
\{x_1, x_6\} = 1, \quad
\{x_2, x_4\} = 1, \quad
\{x_3, x_5\} = 1, \quad
\{x_3, x_6\} = x_2, \quad
\{x_4, x_6\} = x_3 - x_5, \quad
\{x_5, x_6\} = x_2.
\end{equation}

The dynamical functions obtained from \cite{JAF} are
\begin{align}
Q_1 &= (x_3 - x_2 x_1)^2, \\
Q_2 &= x_3 - x_2 x_1, \\
Q_3 &= \exp(x_3 - x_2 x_1), \\
Q_4 &= x_1, \\
Q_5 &= (x_3 - x_2 x_1) x_6, \\
Q_6 &= \sin(x_3 - x_2 x_1).
\end{align}

These satisfy the Poisson bracket \(\{Q_4, Q_5\} = Q_2\), and in general \(\{Q_i, Q_j\} = f_{ij}^k Q_k\), where \(f_{ij}^k\) are the structure constants of the symmetry Lie algebra \(\mathbf{A}_{6,24.v}\). The invariants are \((Q_1, Q_2, Q_4)\) or \((Q_1, Q_2, Q_5)\), and each can serve as the Hamiltonian of an integrable system.

\section {\large {\bf Conclusion}}
We classify all six-dimensional real nilpotent Lie bialgebras of symplectic type and
obtain the Poisson and symplectic
structures on all of the corresponding six-dimensional Poisson-Lie
groups. We also give two examples as physical applications, such that for
these integrable systems the Poisson-Lie group $G$ plays the role of
phase space and its dual Lie group ${\tilde{G}}$ plays the role of
the symmetry group of the system. The calculation of all such systems and
also systems for which the roles of $G$ and ${\tilde{G}}$ are
reversed, and the relation between such systems, are currently under
investigation.\\



\vspace{5mm}

\end{document}